\RequirePackage{fix-cm}
\documentclass[smallextended,natbib]{svjour3}       
\smartqed  
\usepackage{graphicx}
\usepackage{caption}
\usepackage[caption=false]{subfig}
\usepackage{amsfonts}
\usepackage{amsmath}
\usepackage{multirow}
\usepackage{amssymb}
\usepackage{multirow}
\usepackage[colorinlistoftodos, textwidth=4cm, shadow]{todonotes}

\begin{document}

\title{Using Screenshot Attachments in Issue Reports for Triaging}

\titlerunning{Improving Automated Issue Triage with Attached Screenshots}        

\author{Ethem Utku Aktas         \and
        Cemal Yilmaz 
}


\institute{Ethem Utku Aktas \at
              Softtech Inc., Research and Development Center, \\ 
              34947 Istanbul, Turkey \\
              \email{utku.aktas@softtech.com.tr}           
           \and
           Cemal Yilmaz \at
              Faculty of Engineering and Natural Sciences, \\
              Sabanci University, \\ 
              34956 Istanbul, Turkey \\
              \email{cyilmaz@sabanciuniv.edu}           
}

\date{Received: date / Accepted: date}

\maketitle

\begin{abstract}

In previous work, we deployed IssueTAG, which uses the texts present in the one-line summary and the description fields of the issue reports to automatically assign them to the stakeholders, who are responsible for resolving the reported issues. Since its deployment on $January$ $12$, $2018$ at Softtech, i.e.,  the software subsidiary of the largest private bank in Turkey, IssueTAG has made a total of $301,752$ assignments (as of $November$ $2021$). One observation we make is that a large fraction of the issue reports submitted to Softtech has screenshot attachments and, in the presence of such attachments, the reports often convey less information in their one-line summary and the description fields, which tends to reduce the assignment accuracy. In this work, we use the screenshot attachments as an additional source of information to further improve the assignment accuracy, which (to the best of our knowledge) has not been studied before in this context. In particular, we develop a number of multi-source (using both the issue reports and the screenshot attachments) and single-source assignment models (using either the issue reports or the screenshot attachments) and empirically evaluate them on real issue reports. In the experiments, compared to the currently deployed single-source model in the field, the best multi-source model developed in this work, significantly (both in the practical and statistical sense) improved the assignment accuracy for the issue reports with screenshot attachments from $0.843$ to $0.858$ at acceptable overhead costs – a result strongly supporting our basic hypothesis.


\end{abstract}
 
\section{Introduction}
\label{intro}


Issue assignment is the process of assigning the issue reports (also known as the {\em bug reports} or {\em problem reports}) to the stakeholders, who are responsible for resolving the reported issues. As this process is costly, tedious, and error-prone, automating it is of great practical importance, especially for the companies, which receive a large number of issue reports regularly from the field~\citep{jonsson2016automated, lee2017applying, chen2019empirical}.

Softtech\footnote{https://softtech.com.tr}, which constitutes the industrial setup in this work, is one such company. Being a subsidiary of IsBank\footnote{https://www.isbank.com.tr} -- the largest private bank in Turkey, Softtech receives an average of $350$ issue reports from the field on a daily basis for its $400$+ software products comprised of around $100$ millions of lines of code (as of $Nov$ $01$, $2021$). Since these issue reports are typically concerned with business-critical systems, they often need to be handled with utmost importance and urgency. To this end, Softtech and IsBank employ a total of $80$ full-time employees, the sole purpose of which is to carry out the issue triaging process~\citep{aktas2020automated}. Even with this dedicated team of employees, the issue assignment process at Softtech was still suffering due to a number of factors, including the ineffectiveness of maintaining a knowledge base regarding the stakeholders and their responsibilities in an ad hoc manner (to help with the assignments), the ``cost'' of training new triagers, the inevitable friction between the triagers and the development teams in the presence of incorrect assignments, and all of the associated inefficiencies in the triaging process, such as increased turnaround time for resolutions.

To overcome these shortcomings, we, in a previous work, developed an automated issue assignment system, called {\em IssueTAG}, and deployed it at Softtech~\citep{aktas2020automated}. At a very high level, IssueTAG uses the natural language sentences present in the one-line summary and the description fields of the issue reports to assign the reported issues to the development teams (Section~\ref{issuetag}).

Since its deployment on Jan $12$, $2018$, IssueTAG has been making all the assignments in a fully automated manner (about $301,752$ assignments as of $Nov$ $27$, $2021$). Although the assignment accuracy of the system has been slightly lower than that of the human triagers ($0.83$ vs. $0.86$~\citep{aktas2020automated}), this does not prevent the stakeholders from perceiving the deployment system as useful. This is also apparent from a survey we carried out where $79$\% of the participants ``agreed'' or ``strongly agreed'' that IssueTAG is useful~\citep{aktas2020automated}. One reason behind this is that IssueTAG helps the stakeholders defer the responsibility of making the assignments, which is a quite tedious and cumbersome task to carry out manually~\citep{aktas2020automated}. Another reason is that IssueTAG (together with all the modifications made to the triaging process around it) reduces the manual effort required for the assignments by about $5$ person-months per year and improves the turnaround time for resolutions by about $20$\%, on average~\citep{aktas2020automated}.

We have nevertheless been working on further improving the assignment accuracy of the system, especially on figuring out the potential causes of the differences between the accuracy of the human triagers and that of the deployed system. To this end, one observation we make is that a majority of the issue reports submitted to Softtech ($68$\%) have attachments and a majority of these attachments ($84.3$\%) are the actual snapshots of the screens, on which the failures are observed. Although these attachments convey valuable information for issue assignment, they are completely ignored by IssueTAG. As a matter of fact, we are not aware of any work, which utilizes the screenshot attachments in the issue reports for assignment.

Interestingly enough, we also observe that the issue reports with attachments tend to have lower assignment accuracy, compared to those without any attachments ($0.80$ vs. $0.88$, see Section~\ref{motivation} for more information). An in-depth analysis revealed that this could be because the issue reports with the attachments tend to convey less information in their one-line summaries and descriptions as much of the information is already included in the attachments. We, in a study, indeed observed that while the issue reports with attachments had an average of $29$ words, those without any attachments had $41$ words~\citep{aktas2020automated}. 

In this work, we develop and empirically evaluate a number of machine learning approaches, including the multimodal ones, which use the screenshot attachments in issue reports as an additional source of information for assignments. 

In previous work (a poster paper)~\citep{aktas2020exploratory}, we briefly discussed the plausibility of the general idea and presented some preliminary results. In this work, on the other hand, we study the nature of the information present in screenshot attachments in an industrial setup; present a number of additional single-source (utilizing either the textual information or the screenshot attachments present in the issue reports) and multi-source (utilizing both the textual information and the screenshot attachments present in the issue reports) approaches for issue assignment; empirically compare the proposed approaches to a number of alternative approaches (i.e., comparing the multi-source approaches to the single-source approaches); and rigorously evaluate all of the presented approaches by using real issue reports.

More specifically, we address the following research questions in this work:

\begin{list}{-}{}

\item RQ1: What is the status quo in terms of the use of attachments in the issue reports at Softtech?

\item RQ2: How can the screenshot attachments in issue reports be used to further improve the accuracy of the assignments?

\item RQ3: How does taking the screenshot attachments into account affect the overall performance of the system in terms of the training and the prediction times?

\end{list}

The results of our empirical studies conducted on real issue reports submitted to Softtech, strongly suggest that using screenshot attachments as an additional source of information can significantly (both in the practical and statistical sense) improve the accuracy of the assignments at an acceptable cost. In particular, compared to the currently deployed single-source model in the field, the best multi-source model developed in this work, significantly (both in the practical and statistical sense) improved the assignment accuracy for the issue reports with screenshot attachments from $0.843$ to $0.858$ while increasing the training and response (per issue report) times from $190.4$ to $317.2$ seconds and from $0.9$ to $2.17$ seconds, on average, respectively, both of which were in the range of acceptable overheads for Softtech.

The remainder of the paper is organized as follows: Section~\ref{issuetag} provides background information on the previously deployed IssueTAG system; Section~\ref{motivation} analyzes the status quo in terms of the use of attachments in issue reports at Softtech; Section~\ref{feasibility} carries out a feasibility study to better understand the nature of the information present in the screenshot attachments; Section~\ref{method} presents the proposed approaches; Section~\ref{experiments} presents the experiments we carried out to evaluate the proposed approaches; Section~\ref{threats} discusses threats to validity; Section~\ref{relatedWork} summarizes the related work; and Section~\ref{conclusion} concludes with some future work ideas.

\section{IssueTAG}
\label{issuetag}

Softtech receives an average of $350$ software-related issue reports on a daily basis from the field. The reported issues include both the bank clerks having software failures and the bank customers facing software-related problems in any of the banking channels, including mobile, Web, and ATM. Each issue report contains a one-line summary, which captures the essence of the reported issue, and a description, which provides further information regarding the steps for reproducing the reported issues, expected behavior, and observed behavior. Both fields accept natural language sentences in Turkish. In the remainder of the paper, the content of these fields will be referred to as the {\em textual information present in the issue reports}.

Not all reported issues require to change the codebase. Some issues, for example, are resolved by making changes in the databases. In either case, the reported issues, as they typically concern business-critical systems, need to be addressed with utmost importance and urgency.

To carry out the triaging process, two dedicated teams of $80$ full-time employees are employed; IT Help Desk (IT-HD) and Application Support Team (AST). The IT-HD clerks, being consisted of $50$ non-technical personnel, are the first team receiving the issue reports from the field. If they cannot resolve the reported issues by following some basic troubleshooting guidelines, they dispatch them to the proper units at IsBank and Softtech. In the case of software-related issues, the reports are dispatched to the AST team -- a group of $30$ somewhat technical personnel. The AST members, being embedded in software development teams, are capable of resolving most of the issues that do not require to make any modifications in the codebase. If changes in the codebase are needed, then the reports are addressed by the software engineers. 

Before the deployment of IssueTAG, IT-HD clerks were responsible for assigning the issue reports to the software development teams, who are responsible for resolving the reported issues. To this end, IT-HD clerks were using their experiences together with a keyword-based knowledge base, which they collectively maintained in an ad hoc manner. In the case of an incorrect assignment, the issue reports were returned to the IT-HD clerks for reassignment. This was, however, giving rise to issue tossing between the IT-HD clerks and the development teams, causing waste of time.

Note that Softtech prefers to designate development teams as the assignees, so that the dynamic factors, which are quite difficult to take into account during the assignment process, such as the current workloads of the individual developers, the changes in the team structures, and the current status of the  developers (e.g., developers, who are currently on leave of absence), can be addressed within the development teams.

Since the deployment of IssueTAG on $January$ $12$, $2018$ at Softtech, all of the assignments have been made automatically by the system, a total of $301,752$ assignments (as of $November$ $2021$). At a very high level, IssueTAG casts the problem of issue assignment to a classification problem, which takes as input the natural language descriptions present in the one-line summary and description fields of the issue reports and produces as output the assignments~\citep{aktas2020automated}. 

IssueTAG also generates human readable explanations for the assignments, which can be interpreted even by non-technical stakeholders. This was indeed an actual need we discovered only after deploying IssueTAG; the development teams, especially for the incorrect assignments (as this may have an adverse effect on the score cards of the teams), tend to demand explanations as to why the assignments are made in the way they are. 

Another feature implemented by IssueTAG, which is quite important for an automated assignment system operating in a business-critical environment, is a self-monitoring mechanism. In particular, IssueTAG monitors the accuracy of its predictions on a daily basis (by using a change point detection algorithm~\citep{truong2018ruptures, truong2018selective}) and re-train the classification models when the assignment accuracy starts to deteriorate.

\section{Motivation}
\label{motivation}

We have been working on further improving the assignment accuracy of IssueTAG ever since its deployment. To this end, one observation we make is that although a majority of the issue reports submitted to Softtech have attachments, conveying valuable information that can be used toward improving the assignment accuracy, these attachments are completely ignored by IssueTAG. 

To analyze the existing state of affairs in detail, thus to address our first research question  {\em RQ1: What is the status quo in terms of the use of attachments in the issue reports at Softtech?}, we carried out a study. 

In the study, we used a total of $41,042$ real issue reports, which were submitted during the months of March-August in $2019$. In the remainder of the paper, these reports will be referred to as the {\em study data}. We made sure that all of the reports in the study data were actually closed with the ``resolved'' status, indicating that the reported issues were validated and fixed, and that the last assignee for the report (i.e., the one closing the report) is the correct assignee. Note that since the number of issue reports resolved by a development team is a key performance indicator at Softtech, the developers pay utmost attention to correctly indicate the teams closing the issue reports (thus, the correct assignees for the reports).

\begin{figure}[h]
  \centering
  \includegraphics[width=\linewidth]{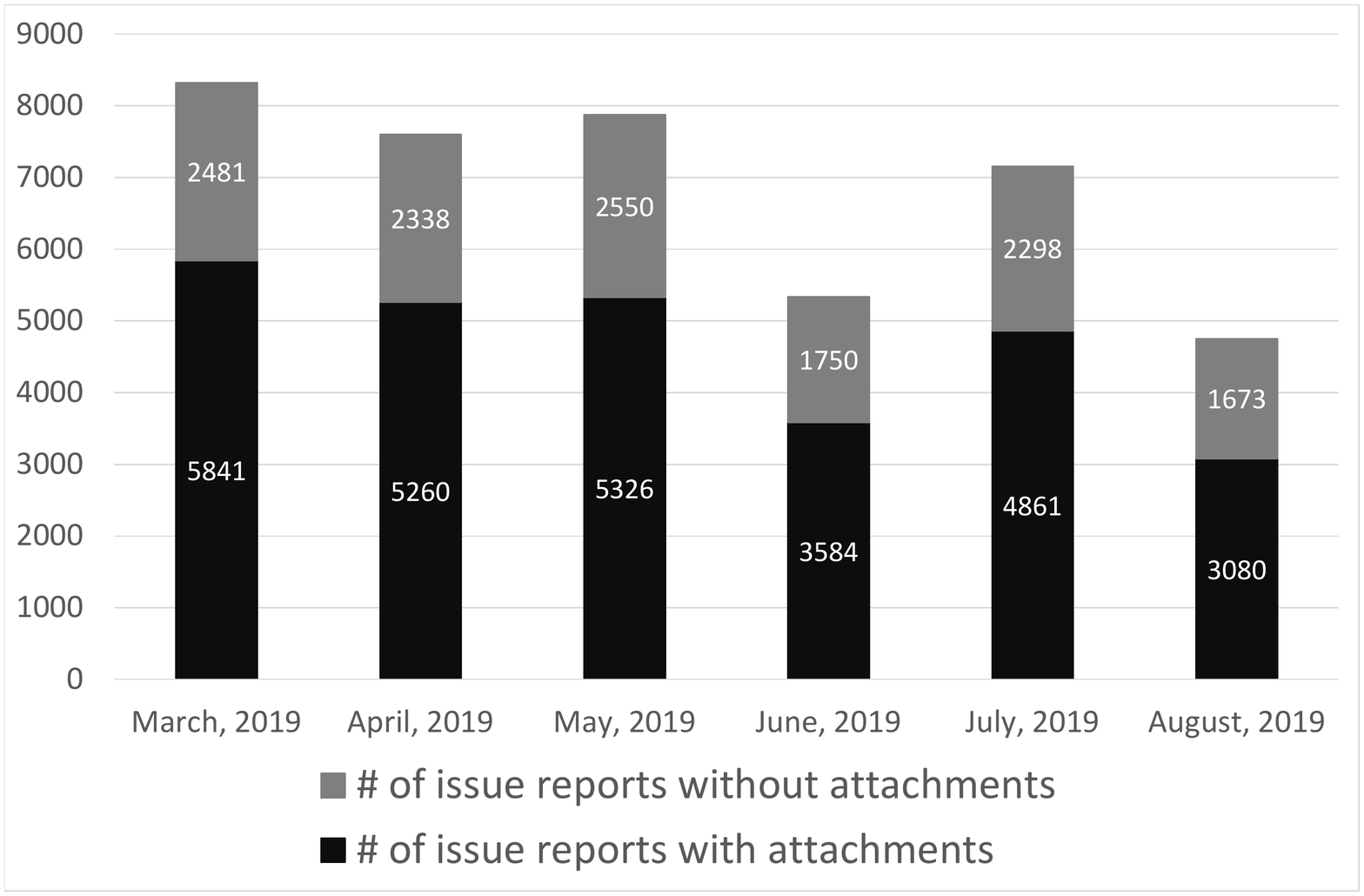}
  \caption{The distribution of the issue reports with and without attachments.}
  \label{stat1}
\end{figure}

We first observed that about $68$\% ($27,952$ out of $41,042$) of all the issue reports had at least one attachment and that the total number of attachments was $34,647$. Figure~\ref{stat1} presents the summary statistics. In particular, $70$\% (out of $8,322$), $69$\% (out of $7,598$), $68$\% (out of $7,876$), $67$\% (out of $5,334$), $68$\% (out of $7,159$), and $65$\% (out of $4,753$) of the issue reports submitted in the months of March-August, respectively, had attachments.

We next observed that although the attachments were of variety of different types (including {\texttt .png},  {\texttt .doc/docx}, {\texttt .xls/xlsx}, {\texttt .msg}, {\texttt .txt}, {\texttt .pdf}, {\texttt .htm/.html}, {\texttt .xml}, and {\texttt .sql}), the most frequently appearing type of attachments was screenshots, capturing the image of the screens, on which the issues were encountered. In particular, among all the issue reports with attachments, $84.30\%$ of them had screenshot attachments; $83.70$\%, $83.25$\%, $84.77$\%, $84.12$\%, $85.89$\%, and $84.06$\% for the months of March-August, respectively. 

We then observed that the issue reports with attachments received significantly lower assignment accuracies, compared to those without any attachments. 

\begin{figure}[h]
  \centering
  \includegraphics[width=\linewidth]{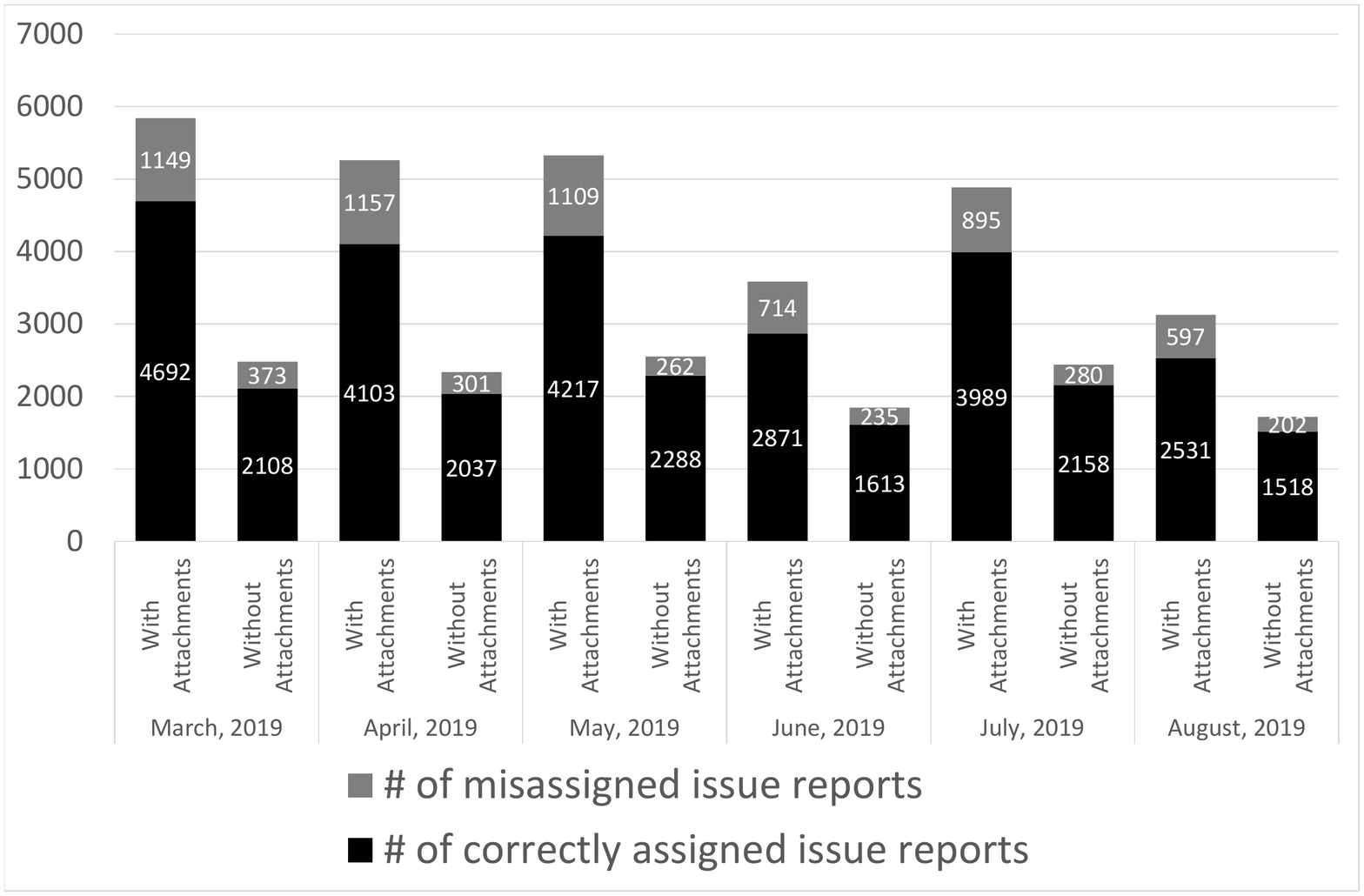}
  \caption{Comparison of reassignments for the issue reports with and without attachments. The average accuracy obtained for the former was $0.80$, that obtained for the latter was $0.88$.}
  \label{stat2}
\end{figure}

More specifically, while the average assignment accuracy for the issue reports with attachments was $0.80$, that for the ones without any attachments was $0.88$ (Figure~\ref{stat2}). And, the monthly assignment accuracies for the former were $0.80$, $0.78$, $0.79$, $0.80$, $0.82$, and $0.81$ for the months of March-August, respectively, whereas those for the latter were $0.85$, $0.87$, $0.90$, $0.87$, $0.89$, and $0.88$.

An in-depth analysis revealed that one potential reason for this is that in the presence of attachments, the issue reports tend to convey less information as much of the information is already included in the attachments. This phenomenon is indeed also apparent from the number of words included in the issue reports with and without attachments. For example, while the average number of words in the issue reports with screenshot attachments is $29$, that of the reports without any attachments is $41$.

\begin{table*}
  \caption{Example issue reports with screenshot attachments}
  \label{tab:sample_bug}
  \resizebox{\columnwidth}{!}
  {\begin{tabular}{ccl}
    \hline\noalign{\smallskip}
    issue & one-line summary & description\\
    \noalign{\smallskip}\hline\noalign{\smallskip}
    
    1 & $[ScreenCode]$ & The screen $[ScreenCode]$ does not appear at all terminals in branch $[BranchCode]$; the error \\
    &&given in the attachment is observed.\\


    2 & $[ErrorCode]$ & Although the requested limits have been updated, we receive the attached error during \\
    && the approval process of the customer's ($[CustomerCode]$) request.\\

    3 & $[ScreenCode]$/$[TransactionCode]$ & We receive the attached error for the transaction $[TransactionCode]$ on the screen $[ScreenCode]$. \\
    \noalign{\smallskip}\hline
  \end{tabular}}
\end{table*}

In this work, since the screenshot attachments are the most frequently appearing type of attachments at Softtech and since there is still room for improving their assignment accuracies, we opted to solely focus on the screenshot attachments. We, in particular, conjecture that using screenshot attachment as an additional source of information (i.e., together with the textual information present in the issue reports), can improve the accuracy of the assignments.

\section{Feasibility Study}
\label{feasibility}

To better understand the nature of the information conveyed in screenshot attachments, we first carried out a feasibility study. The results of this study were indeed instrumental in designing the solution approaches introduced in Section~\ref{method}.

To carry out the feasibility study, we have manually analyzed a number of issue reports with screenshot attachments, which were incorrectly assigned by IssueTAG. Table~\ref{tab:sample_bug} presents some examples, which we will use to summarize the insights we gained throughout the study. Note that due to certain security and privacy concerns, the table provides only the one-line summaries and the descriptions for the aforementioned issue reports where the actual error and transaction codes are obscured and the actual screenshots are omitted.

Regarding the first issue report (Table~\ref{tab:sample_bug}), one would expect that the screen code indicated in both the summary and the description fields of the report would be instrumental in assigning the report. It, however, turns out that this screen code has never occurred in any of the historical issue reports, which simply renders the natural language descriptions present in the report useless. 

When we manually analyzed the screenshot attachment in the report, we, to our surprise, observed that the error message mentioned in the description was a generic ``HTTP 404 - Web page cannot be found'' error, which is, indeed, not useful at all either. On the other hand, the textual information present in the remainder of the screen, such as, the titles of the open tabs, clearly indicated that the error message was indeed emitted by the retail loan management module. Had the text been extracted from the attached screenshot and used for the assignment, the report would have been assigned to the correct development team.

Regarding the second issue report (Table~\ref{tab:sample_bug}), although,  at a first glance, this report seems to be quite similar to the first report in the sense that both reports have a screenshot of the error message emitted as an attachment, a manual analysis of the attachment revealed some interesting differences. 

More specifically, the image attached to the first report was the screenshot of a screen created by the software module responsible for the failure. The image attached to the second report, on the other hand, was a screenshot obtained from a general-purpose workflow engine, visualizing the business process, in which the failure was observed. That is, the workflow engine was not responsible for the failure. Instead, the failure was caused by a module responsible for handling one of the tasks in the visualized workflow. That is, the screenshot attachment alone was not enough for the assignment. More specifically, the textual information present both in the report and in the screenshot attachment should have been used together to correctly assign the report as combining both sources of information  indicated that the failure was related to a module handling credit card limit operations.

Regarding the third issue report (Table~\ref{tab:sample_bug}), interestingly enough, the image of the screen attached to this report has quite a different look and feel, compared to the images of the screens attached to the first two reports. 

It turns out that the screenshot in this report comes from a module written in COBOL programming language running on mainframes, whereas the other screenshots were images of some web-based screens created by using recent web technologies. Although this suggests that classifying the screen images (by using image classification) can help improve the assignment accuracy, an in-depth analysis quickly revealed that this may not be the case in practice (at least for Softtech). The reason is two folds. First, the different development teams at Softtech use the same graphical user interface (GUI) frameworks (we identified three such frameworks including the one used on the mainframes) with the same (or similar) strict GUI design guidelines. Therefore, the look and feel of the screens produced by different development teams are typically quite similar to each other (if not the same). Second, it is not unusual for a development team to use multiple GUI frameworks in their products. For example, a team can use a web-based GUI framework for the non-technical end users and a mainframe-based GUI framework for the more technical users. This, however, did not prevent us from experimenting with the single-source models based on the image classification of the screenshot attachments, which we indeed used as a baseline (Section~\ref{method}). On the other hand, we clearly observed that, as was the case with the first two issue reports,  extracting the textual information from the screenshot attachment in the third issue report would again help us correctly assign the report as the text appearing on the screen indicated that the reported failure was related to a module handling the cheque transactions.

\section{Approach}
\label{method}

\begin{figure}[h]
  \centering
  \includegraphics[width=\linewidth]{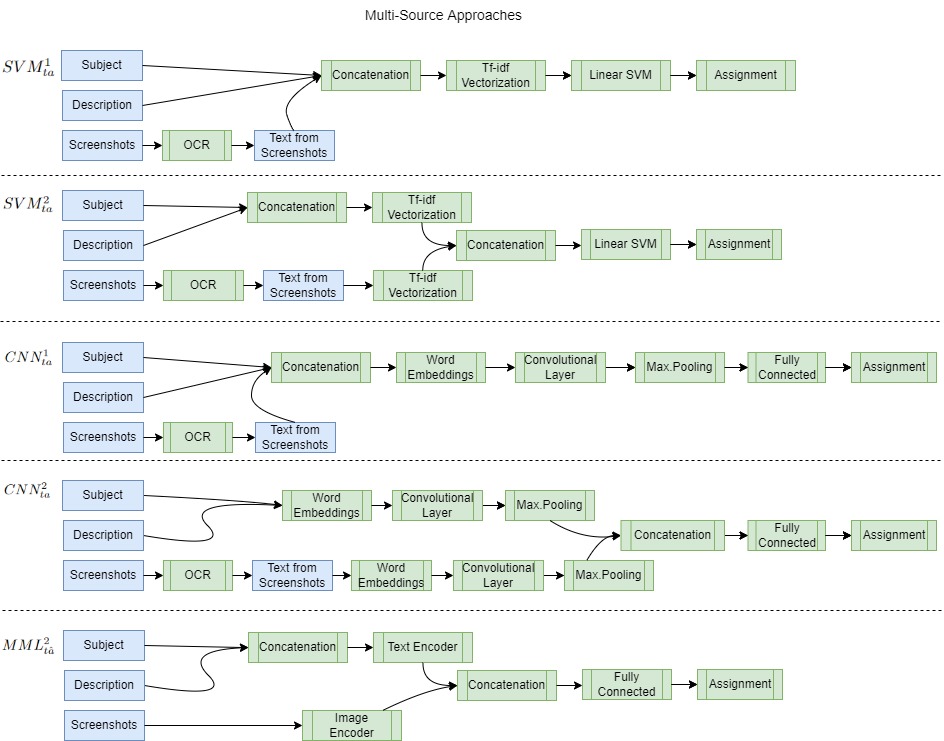}
  \caption{Proposed multi-source approaches.}
  \label{fig:multi_approaches}
\end{figure}

\begin{figure}[h]
  \centering
  \includegraphics[width=\linewidth]{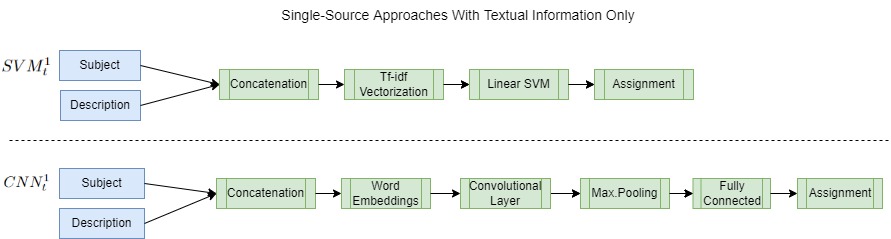}
  \caption{Proposed single-source approaches using textual information only.}
  \label{fig:single_approaches_text}
\end{figure}

\begin{figure}[h]
  \centering
  \includegraphics[width=\linewidth]{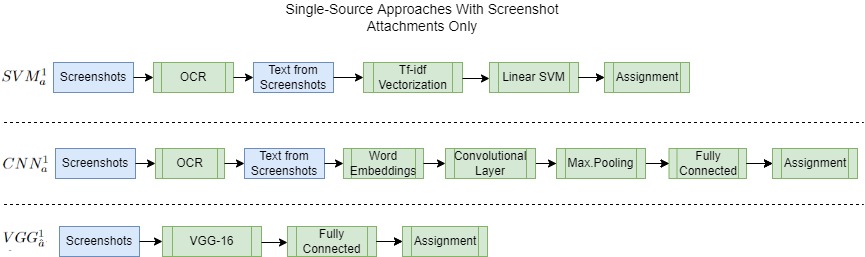}
  \caption{Proposed single-source approaches using screenshots only.}
  \label{fig:single_approaches_image}
\end{figure}

With all the insights we gained from our manual analysis in mind, we have developed a number of approaches to take the screenshot attachments into account when assigning the issue reports. Figures~\ref{fig:multi_approaches},~\ref{fig:single_approaches_text} and~\ref{fig:single_approaches_image} summarize these approaches.

One commonality between these approaches is that they all cast the problem at hand to a classification problem where the class labels to be predicted represent the development teams, to which the issue reports should be assigned. Furthermore, the approaches, which analyze the text extracted from the one-line summaries, descriptions, and/or screenshot attachments, pre-process the text before any analysis. In particular, we first tokenize the words in the extracted text, then eliminate the special characters, such as symbols and punctuation characters, and finally remove the stop-words, which do not contribute to the assignments at all. We then use n-grams to capture the vocabulary in issue reports, where a word or group of n-words are represented with a numerical value which depicts the importance of the term for that issue report. 


The proposed approaches, however, differ from each other in the sources of information they use and in the way they model the classification problem. At a very high level, they can be grouped into three broad categories: {\em multi-source}, {\em single-source with textual information only} (for short, {\em single-source-report}), and {\em single-source with screenshot attachments only} (for short, {\em single-source-attachment}). The multi-source approaches utilize both the textual information and the screenshot attachments present in the issue reports, whereas the single-source models utilize either the textual information in the issue reports or the screenshot attachments (but not both). As the single-source approaches evaluate the individual contributions of the textual information and the screenshot attachments in the issue reports separately, they serve as the baselines when the claims of this work are considered.

Next, we, therefore, introduce the multi-source approaches first and then discuss the single-source approaches.

\subsection{Multi-Source Approaches}
\label{multiSource}

In multi-source approaches, we leverage both the screenshot attachments and the textual information present in the issue reports, i.e., the text in the one-line summaries and the  descriptions of the reports. We, in particular, develop five different approaches, namely $SVM_{ta}^{1}$, $SVM_{ta}^{2}$, $CNN_{ta}^{1}$, $CNN_{ta}^{2}$, and $MML_{t\hat{a}}^2$. 

Note that while the names of the approaches provide clues about the classification models used, the sub-scripted symbols and the super-scripted numbers indicate the sources of information leveraged and the number of channels used in the models, respectively. In particular, the sub-scripts $t$ and $a$ indicate the inclusion of the textual information present in the issue reports and the inclusion of the screenshot attachments in modeling, respectively. Furthermore, the presence of a hat above $a$ (i.e., $\hat{a}$) specifies that the screenshots are processed as images (i.e., visual features are used in the models), whereas the absence of the hat indicates that only the textual information extracted from the screenshots are processed. And, the number of channels indicate whether the features extracted from $t$ and $a$ are merged together (indicated by the super-script $1$) or treated separately (indicated by the super-script $2$).

\subsubsection{$SVM_{ta}^1$}
\label{svmta1}

In this approach, we first extract the textual information present in the screenshot attachments using optical character recognition (OCR)~\citep{smith2007overview} and then merge it (i.e., into a single channel) with the textual information present in the one-line summary and the description fields of the issue reports (Figure~\ref{fig:multi_approaches}).

More specifically, we represent each issue report as a vector in a multi-dimensional space by using the bag of words (BoW) model with the well-known $tf$-$idf$ scoring scheme~\citep{manning2008introduction}. Each element in the vectorized form of an issue report, represents a term and the value of the element (i.e., the $tf$-$idf$ score of the respective term) depicts the importance of the term for the issue report. The more a term appears in an issue report (i.e., the higher the {\em term frequency} score $tf$) and the less it appears in other issue reports (i.e., the higher the {\em inverse document frequency} score $idf$), the more important the term becomes for the report (i.e., the higher the $tf$-$idf$ score, thus the weight, of the term is).

To train the classification models, we feed the $tf$-$idf$ representations of the issue reports to a linear SVM model~\citep{pedregosa2011scikit}. We, in particular, opted to use the linear SVM models with the BoW representations, because the results of our earlier studies strongly suggest that these models offer us the best prediction accuracy in our industrial setup with manageable training and prediction costs (in terms of the training and prediction times required as well as the amount of training data needed)~\citep{aktas2020automated}. 

This is, indeed, the model that has been used by IssueTAG in the field since its deployment~\citep{aktas2020automated} (Section~\ref{issuetag}). The difference is that while the deployed system leverages only the textual information present in the issue reports and ignores the attachments, $SVM_{ta}^1$ leverages both sources of information.



Throughout the paper, we train the SVM models by using $scikit-learn$~\citep{pedregosa2011scikit} with a linear kernel and extract the text from the screenshots by using $py-tesseract$~\citep{smith2007overview}.


\subsubsection{$SVM_{ta}^2$}
\label{svmta2}

One observation we make regarding the text extracted from the screenshots and the text present in the one-line summaries and the descriptions, is that often exhibit different properties. More specifically, the latter is typically written using a formal language with little or no language errors (e.g., typos and grammar mistakes) at all. The former, on the other, typically has many typos due to the OCR errors. Interestingly enough, we also observe that OCR tends to repeatedly make the same or similar mistakes. For example, the same sequence of characters tend to be recognized wrongly in exactly the same manner.

To account for these differences, we have developed a multimodal classification approach ($SVM_{ta}^2$) by treating the text extracted from the issue reports and text extracted from the screenshot attachments as two different channels (Figure~\ref{fig:multi_approaches}). More specifically, while the combined text obtained from the one-line summaries and the descriptions forms a channel, the text extracted by using OCR forms another channel. 

In this approach, although we encode the information flowing through each channel by using the BoW model with the $tf$-$idf$ scoring scheme (as explained in Section \ref{svmta1}), we compute the $tf$-$idf$ scores on a per channel basis. That is, the term frequencies and the inverse document frequencies are computed separately for each channel. Therefore, given an issue report with a screenshot attachment, we compute two vectors (one per channel), which are then appended to each other before being fed to a linear SVM model.

\subsubsection{$CNN_{ta}^1$ and $CNN_{ta}^2$}
\label{cnnta123}

The BoW models we used in the first two approaches (Sections~\ref{svmta1} and~\ref{svmta2}), do not necessarily take the contexts of the terms appearing in the issue reports into account when making the assignments. To overcome this issue, we, in this section, use deep neural networks to generate word embeddings and use them for the assignments~\citep{lee2017applying}. In a nutshell, word embeddings are the vectorized forms of the words, such that the vectors (i.e., the embeddings) of the semantically similar (or related) words are close to each other in a multi-dimensional space.

Note that all the issue reports we are dealing with in this work are written in Turkish. Although the language used in these reports are quite formal, the reports include an extensive use of the finance jargon as well as the company jargon, which has been developed over the years with a great deal of abbreviations. We, therefore, chose to train our own word embeddings by using the issue database maintained at Softtech. To this end, we have used the Keras embedding layer~\citep{chollet2015keras}. In particular, the word embeddings were initialized with random weights and fine-tuned throughout the training process.

Once the word embeddings are learnt, we used them to train convolutional neural networks (CNN) -- an approach inspired from \citep{lee2017applying}, which presents a state-of-the-art application of the word embeddings for issue assignment (Figure~\ref{fig:multi_approaches}). More specifically, for the text flowing through a channel (either from the one line-summaries and descriptions or from the screenshot attachments), we first represent it with the word embeddings, conveying the semantics. We then apply a convolution process using a sample-based discretization approach, called {\em max-pooling} \citep{goodfellow2016deep}. Finally, the outputs are concatenated and, through a fully-connected layer and softmax regression, the probabilities for the assignees are computed. To prevent overfitting, we apply dropout as well as L2 regularization. The interested reader can refer to \cite{lee2017applying} for further details. We, in particular, experiment with two different models, namely $CNN_{ta}^1$ and  $CNN_{ta}^2$. The former model uses a single channel, into which the text extracted from the screenshots attachments and from the textual information present in the issue reports are merged. The latter model, on the other hand, uses two channels by treating the text extracted from the screenshots and from the issue reports separately.


\subsubsection{$MML_{t\hat{a}}^2$}
\label{MMLta2}


All of the approaches we have discussed so far (Sections~\ref{svmta1}-\ref{cnnta123}) leverage the textual information extracted from the screenshot attachments by using OCR and completely ignore the visual features. The $MML_{t\hat{a}}^2$ model, on the other hand, uses the visual features extracted from the screenshots together with the textual information present in the one-line summaries and descriptions of the issue reports (Figure~\ref{fig:multi_approaches}).




Modality is the mode in which something is experienced, such as vision, text and audio \citep{baltrusaitis2018multimodal}. With the $MML_{t\hat{a}}^2$ model, we aim to build a basic architecture to integrate visual and textual features of the issue reports for our specific classification task. We use the fastText embeddings~\citep{joulin2016bag, bojanowski2017enriching} to obtain the vector representations for the textual data and the ResNet image recognition model~\citep{he2016deep} for the visual representations. The vector outputs are passed through a linear layer separately to reduce their dimension. Rectified linear activation function (ReLU) is applied on both, that outputs the input directly if it is positive, zero, otherwise. The resulting textual and visual features are combined (or fused) to reduce their dimension, passed through the activation function ReLU, and a dropout is applied, which is a regularization technique to forget some of the information learned by the network. The final representations are passed through a fully-connected layer and softmax function for classification. We use the PyTorch deep learning framework~\citep{paszke2019pytorch} to build the multimodal model. 


\subsection{Single-Source Approaches using Textual Information Only}
\label{singleSourceText}

We use the approaches discussed in this section to evaluate the effect of the textual information present in the issue reports on the assignment accuracy. To this end, the aforementioned approaches use only the one-line summaries and the descriptions of the issue reports, and completely ignore the screenshot attachments.

We, in particular, experiment with two approaches: $SVM_{t}^1$ and $CNN_{t}^1$ (Figure~\ref{fig:single_approaches_text}). In these models, the text present in the one-line summaries and the descriptions of the issue reports are analyzed by using the SVM and CNN models as discussed in Section~\ref{svmta2} and Section~\ref{cnnta123}, respectively.

\subsection{Single Source Approaches using Attachments Only}
\label{singleSourceAttachment}

While the approaches in Section~\ref{singleSourceText} are used to evaluate the amount of information conveyed in the textual descriptions of issue reports, which can be used toward the assignments, the approaches we study in this section carry out the same analysis for the information conveyed in the screenshot attachments. To this end, we use only the screenshot attachments for assigning the issue reports to the stakeholders and completely ignore the one-line summaries and the descriptions of the reports. Note that the approaches we present both in this section and in the previous section (Section~\ref{singleSourceText}) also serve as a baseline for the multi-source approaches introduced in Section~\ref{multiSource}.


More specifically, we experiment with $3$ approaches: $SVM_{a}^1$, $CNN_{a}^1$, and $VGG_{\hat{a}}^1$. The first two approaches extract the textual information present in the attachments using OCR and use the extracted text to train the SVM and CNN models in exactly the same manner discussed in Section~\ref{svmta2} and Section~\ref{cnnta123}, respectively.


The last approach ($VGG_{\hat{a}}^1$), on the other hand, uses the visual features (rather than the textual features) extracted from the screenshot attachments. In particular, we use the well-known VGG-16 architecture~\citep{simonyan2014very} implemented by Keras~\citep{chollet2015keras}, which includes five convolutional blocks consisted of a total of thirteen convolutional layers, followed by three fully connected layers. We use transfer learning, in other words, we fix the weights of all of the convolutional layers during training, replace the last fully connected layers with new fully connected layers, and train only the new layers for assigning the issue reports. The first layer after the convolutional layers flattens the input vector to obtain a one dimensional vector, then a dense layer is used to reduce the dimension of the vector where a ReLU function is applied, and finally a fully connected layer and softmax function is used for classification. 

\subsection{Hybrid Approach}
\label{hybridApproach}

One observation we made in the experiments was that although the multi-source approaches improve the assignment accuracy for the issue reports with screenshot attachments, they tend to slightly reduce the accuracy for the issue reports without any attachments. We believe that this was because having no information flowing through the respective channel in the absence of any attachments tend to make the issue reports close to each other due the aforementioned commonality. 

We, therefore, also develop a hybrid approach, called $SVM_{hybrid}$, by combining the best performing multi-source model in the experiments, i.e., $SVM_{ta}^{2}$, together with the best performing single-source model, i.e., $SVM_{t}^{1}$ (Section~\ref{experiments}). More specifically, we use the $SVM_{ta}^{2}$ model for assigning the issue reports with screenshot attachments and the $SVM_{t}^{1}$ model for assigning the ones without any attachments. 

\section{Experiments}
\label{experiments}

To evaluate the proposed approaches, we have carried out a series of experiments.

\subsection{Subject Issue Reports}
\label{subjects}

In the experiments, we used the real issue reports submitted to Softtech. Table~\ref{tab:data} presents the summary statistics for these issue reports.

\begin{table*}
  \caption{Summary statistics for the issue reports used in the experiments.}
  \label{tab:data}
  \resizebox{\columnwidth}{!}
  {\begin{tabular}{|l|r|r|r|r|r|}
    \hline
    \multicolumn{1}{|c|}{month
}& \multicolumn{1}{|c|}{issue reports} & \multicolumn{1}{|c|}{issue reports} & \multicolumn{1}{|c|}{issue reports} & & \multicolumn{1}{|c|}{distinct}\\
     \multicolumn{1}{|c|}{of creation} & \multicolumn{1}{|c|}{with screenshots} & \multicolumn{1}{|c|}{with attachments} & \multicolumn{1}{|c|}{without attachments} & \multicolumn{1}{|c|}{total} & \multicolumn{1}{|c|}{assignees} \\
     \hline
    August, 2019 & 2,589 & 3,080 & 1,673 & 4,753 & 49\\
    July, 2019 & 4,175 & 4,861 & 2,298 & 7,159 & 53\\
    June, 2019 & 3,015 & 3,584 & 1,750 & 5,334 & 51\\
    May, 2019 & 4,515 & 5,326 & 2,550 & 7,876 & 53\\
    April, 2019 & 4,379 & 5,260 & 2,338 & 7,598 & 58 \\
    March, 2019 & 4,889 & 5,841 & 2,481 & 8,322 & 52\\
    \hline
    initial set total & 23,562 & 27,952 & 13,090 & 41,042 & 63\\
    \hline
    February, 2019 & 3,838 & 4,890 & 2,058 & 6,948 & 52\\
    January, 2019 & 4,951 & 6,180 & 2,700 & 8,880 & 49\\
    December, 2018 & 3,741 & 4,316 & 1,945 & 6,261 & 47\\
    November, 2018 & 4,349 & 5,065 & 2,293 & 7,358 & 49\\
    October, 2018 & 3,976 & 4,688 & 2,368 & 7,056 & 49\\
    September, 2018 & 4,196 & 5,052 & 2,375 & 7,427 & 49\\
    \hline
    additional set total & 25,051 & 30,191 & 13,739 & 43,930 & 60\\
    \hline
    grand total & 48,613 & 58,143 & 26,829 & 84,972 & 68\\
    \hline
  \end{tabular}}
\end{table*}

For the initial set of experiments, where the goal was to evaluate all the proposed approaches introduced in Section~\ref{method}, we used a total of $41,042$ issue reports submitted to $63$ distinct development teams between the months of March and August in $2019$ (Table~\ref{tab:data}). In particular, we utilized the issue reports submitted in August as the test set and all the remaining issue reports submitted from March to July as the training set. 

After determining the best performing multi-source and single-source approaches, we performed a series of statistical significance tests to figure out whether the differences between these approaches are statistically meaningful. To this end, we used an additional $43,930$ issue reports submitted to $60$ distinct teams between September, 2018 and February, 2019 (Table~\ref{tab:data}).

We made sure that all of the issue reports used in the analyses (i.e., the ones mentioned above) were closed with the ``resolved'' status. This guaranteed that all of the selected reports actually indicated real issues and that the development teams closing the reports were the correct assignees for the respective reports. 

Furthermore, for each issue report, we have first figured out whether the report had any attachments or not. In the presence of any screenshot attachments, which was determined by examining the file extensions and attachment types, we have extracted them for latter processing. 


\subsection{Evaluation Framework}
\label{evaluation}

All told, we used a total of $84,972$ real issue reports submitted to $68$ distinct teams for the evaluations (Table~\ref{tab:data}).

To evaluate the correctness of the assignments (thus, to address our second research question), we have computed both the accuracy and F-measure metrics for the assignments. More specifically, the accuracy (A) was computed as the ratio of the number of correctly assigned issue reports to the total number of issue reports in the test set. And, F-measure (F) was computed as the harmonic mean of the precision (P) and recall (R), giving equal importance to both metrics. For a given development team (i.e., for a given class), the precision of the assignments is computed as the ratio of the number of correctly assigned reports to the team to the total number of issue reports assigned to the team. The recall is, on the other hand, computed as the ratio of the number of correctly assigned reports to the team to the total number of issue reports that should have been assigned to the team. Since multiple classes were present in the experiments, we, in this work, report the weighted values. Note that all of the aforementioned metrics take on a value between $0$ and $1$ inclusive and that the larger the value, the better the assignments, thus the proposed approaches, are.

For the statistical significance tests, we have repeated the experiments $30$ times for each experimental setup by utilizing different training and/or test sets. The results were then analyzed by using the non-parametric Wilcoxon rank sum test~\citep{wilcoxon1992individual} where a p-value less than $0.05$ was considered to be statistically significant.

Furthermore, since this work targets a system operating in a production environment, excessive runtime overheads are simply not acceptable. To evaluate the performance of the proposed approaches (thus, to address our third research question), we also measure the running times of the important tasks. We, in particular, measure the average amount of time required for both training the classification models and using them for predictions as well as the average running times required for extracting the text from screenshot attachments using OCR.

\subsection{Data and Analysis}
\label{data}

We have carried out all the experiments and used the results to address our remaining research questions, namely RQ2 and RQ3. Note that the first research question RQ1 has already been addressed in Section~\ref{motivation}.

\begin{table*}
\small
\caption{Accuracy (A), precision (P), recall (R), and F-measure (F) values obtained from different approaches. Note that the approaches that require the presence of screenshot attachments cannot be evaluated on the issue reports with no attachments.}
\label{tbl:results_all}      
\resizebox{\columnwidth}{!}   
{\begin{tabular}{|l|l|rrrr|rrrr|rrrr|}
\hline
\multirow{2}{*}{approach}&
\multirow{2}{*}{model}&
\multicolumn{4}{|c|}{test data w/o screenshots} & \multicolumn{4}{|c|}{test data w/ screenshots} &\multicolumn{4}{|c|}{all test data}  \\
\cline{3-14}
&  & A & P & R & F & A & P & R & F & A & P & R & F \\
\hline
\multirow{5}{*}{\begin{tabular}{l}multi-source\end{tabular}} & $SVM_{ta}^{1}$ & 0.844 & 0.851 & 0.844 & 0.837 & 0.821 & 0.814 & 0.821 & 0.812 & 0.832 & 0.826 & 0.832 & 0.823 \\
& $SVM_{ta}^{2}$ & 0.848 & 0.855 & 0.848 & 0.840 & 0.858 & 0.851 & 0.858 & 0.848 & 0.854 & 0.850 & 0.854 & 0.846 \\
& $CNN_{ta}^{1}$ & 0.825 & 0.831 & 0.825 & 0.819 & 0.789 & 0.794 & 0.789 & 0.779 & 0.819 & 0.810 & 0.819 & 0.804 \\
& $CNN_{ta}^{2}$ & 0.819 & 0.826 & 0.819 & 0.812 & 0.833 & 0.820 & 0.833 & 0.821 & 0.826 & 0.827 & 0.826 & 0.819 \\
& $MML_{t\hat{a}}^2$ & - & - & - & - & 0.411 & 0.790 & 0.411 & 0.530 & 0.411 & 0.790 & 0.411 & 0.530 \\
\hline
\multirow{2}{*}{\begin{tabular}{l} single-source\\w/ textual information\end{tabular}} & $SVM_{t}^{1}$ & 0.851 & 0.860 & 0.851 & 0.845 & 0.843 & 0.836 & 0.843 & 0.834 & 0.848 & 0.845 & 0.848 & 0.839 \\
& $CNN_{t}^{1}$ & 0.826 & 0.839 & 0.826 & 0.825 & 0.819 & 0.817 & 0.819 & 0.812 & 0.828 & 0.818 & 0.828 & 0.816 \\
\hline
\multirow{3}{*}{\begin{tabular}{l} single-source\\w/ attachments\end{tabular}} &  $SVM_{a}^{1}$ & - & - & - & - & 0.705 & 0.701 & 0.705 & 0.690 & 0.705 & 0.701 & 0.705 & 0.690 \\
& $CNN_{a}^{1}$ & - & - & - & - & 0.696 & 0.712 & 0.696 & 0.684 & 0.696 & 0.712 & 0.696 & 0.684 \\
& $VGG_{\hat{a}}^1$ & - & - & - & - & 0.046 & 0.008 & 0.046 & 0.007 & 0.046 & 0.008 & 0.046 & 0.007 \\
\hline
\multirow{1}{*}{\begin{tabular}{l} hybrid\end{tabular}} & $SVM_{hybrid}$ & 0.851 & 0.860 & 0.851 & 0.845 & 0.858 & 0.851 & 0.858 & 0.848 & 0.855 & 0.850 & 0.855 & 0.846 \\
\hline
\end{tabular}}
\end{table*}

\subsubsection{Regarding RQ2: How can the screenshot attachments in issue reports be used to further improve the accuracy of the assignments?}
\label{rq2}

Table~\ref{tbl:results_all} presents the results we obtained from the experiments we carried out to address RQ2.


{\bf Using the textual vs. visual features in screenshot attachments.} Comparing the single-source models, which use only the screenshot attachments for issue assignment (i.e., the single-source-attachment models), with each other, we first observed that using only the visual features extracted from the screenshot attachments (i.e., $VGG_{\hat{a}}^1$) is not helpful at all. In particular, the accuracy of the assignments obtained from the $VGG_{\hat{a}}^1$ model was $0.046$ (Table~\ref{tbl:results_all}). 

We believe that this was mainly due to the fact that a small number of user interface (UI) frameworks have been used throughout Softtech together with a set of quite strict guidelines regarding the UI designs, including the color palette to use and the general design templates to follow. Consequently, the screens produced by different development teams typically have the same or similar look-and-feel, which makes it quite difficult to distinguish between the producers of these screens by using only the visual features. 

Note that the models that require the presence of screenshot attachments in order to operate, such as $VGG_{\hat{a}}^1$, cannot be evaluated on the issue reports without any attachments; explaining the missing values, i.e., the ``-'' symbols, in Table~\ref{tbl:results_all}. We, therefore, report the accuracy of these models only for the issue reports with screenshot attachments.

We next observed that using the textual features present in the screenshots, compared to using the visual features, were significantly better at making accurate assignments. More specifically, the best accuracy obtained from the single-source-attachment models, i.e., an accuracy of $0.705$ (Table~\ref{tbl:results_all}), was obtained from the $SVM_{a}^{1}$ model, which leverages the text extracted from the screenshots for the assignments. Indeed, these results further support the claims of the paper that the text present in the screenshot attachments convey information, which can be leveraged for issue assignment.

{\bf Regrading the information content of the one-line summaries and descriptions in the issue reports.} We then observed that, even in the presence of screenshot attachments, the textual information present in the one-line summary and the description fields of the issue reports were still quite valuable for the assignments. Among all the single-source models, the best accuracy for the issue reports with screenshot attachments, was still obtained from a single-source-report model, namely the $SVM_{t}^{1}$ model. In particular, the accuracy of the aforementioned model was $0.843$, which was significantly better than the ones obtained from the single-source-attachment models (Table~\ref{tbl:results_all}).

We believe that this was mainly due to the fact that, in the software systems maintained by Softtech, developing a single screen typically requires the involvement of multiple teams. For example, a screen associated with the credit card operations, which is maintained by the credit cards team, can use services in the background, which are developed by the customer information management team and the commercial/individual credit team. Therefore, a failure observed on this screen may be caused from any of these services, which are maintained by different development teams. Consequently, given a screen, without knowing the symptoms of the reported issue, which is typically given in the one-line summary and the description fields of the issue reports, it may not be possible to determine the development team responsible for resolving the reported issue. 

Note that, in the analysis above, we used only the issue reports with screenshot attachments for the comparisons, so that different approaches could fairly be evaluated by using exactly the same set of issue reports. Similarly, since the ultimate goal of our experiments is to evaluate the effect of using the screenshot attachments for issue assignment, in the remainder of the analysis we, unless otherwise stated, focus on the results obtained from the issue reports with screenshot attachments. However, since the proposed approach should not adversely affect the assignment accuracy for the issue reports without any attachments, we also report (Table~\ref{tbl:results_all}) and analyze (later on) the overall accuracy of the proposed approach by using the issue reports both with and without attachments.


{\bf Using both sources of information.} We finally analyzed the results obtained from our multi-source models (Table~\ref{tbl:results_all}). The first thing we observed, which is also well-aligned with our discussion regarding the $VGG_{\hat{a}}^1$ model above, was that using the visual features extracted from the screenshot attachments were not helpful at all. In particular, the assignment accuracy obtained from combining the visual features extracted from the screenshot attachments with the textual information present in the issue reports (i.e., the $MML_{t\hat{a}}^{2}$ model) was $0.411$.

We, however, observed that extracting the text from the screenshot attachments and combining it with the text present in the one-line summary and the description fields of the issue reports (i.e., the $SVM_{ta}^{1}$, $SVM_{ta}^{2}$, $CNN_{ta}^{1}$, and $CNN_{ta}^{2}$ models) profoundly increased the accuracy of the assignments, compared to using the single-source models. While the best accuracy obtained from the former models was $0.858$, the one obtained from the latter models was $0.843$, supporting the claims of the paper (Table~\ref{tbl:results_all}).


We next observed that the SVM models (i.e., $SVM_{ta}^{1}$ and $SVM_{ta}^{2}$) generally performed better than the CNN models (i.e., $CNN_{ta}^{1}$ and $CNN_{ta}^{2}$), a phenomenon we observed a number of times in our previous works when it comes to analyzing the issue report repository of Softtech~\citep{aktas2020turkish}. The best accuracy obtained from the former models was $0.858$, whereas that obtained from the latter models was $0.833$ (Table~\ref{tbl:results_all}).

We then observed that using a multi-modal approach by treating the texts coming from the screenshot attachments and from the issue reports separately performed better than treating them as one single text (Section~\ref{method}). More specifically, while the accuracy of $SVM_{ta}^{2}$ was $0.858$, that of  $SVM_{ta}^{1}$ was $0.821$ (Table~\ref{tbl:results_all}).

{\bf Accuracy of the multi-source models on the issue reports without any screenshot attachments.} When we compared the assignment accuracy of the best performing multi-source model, i.e., $SVM_{ta}^{2}$, to that of the best performing single-source model, i.e., $SVM_{t}^{1}$, for the issue reports without any screenshot attachments, we observed the $SVM_{ta}^{2}$ model slightly reduced the assignment accuracy; $0.851$ vs. $0.848$. We believe that this is because, in the absence of any attachments, having no information flowing through the respective channel in the multi-source models tend to make the issue reports close to each other due to this commonality. 

Using our hybrid model $SVM_{hybrid}$, on the other hand, resolved this issue. Although the $SVM_{hybrid}$ model provided a similar overall accuracy with the $SVM_{ta}^{2}$ model ($0.855$ vs. $0.854$), the former prevented the assignment accuracy of the issue reports without any screenshot attachments from suffering (Table~\ref{tbl:results_all}). In particular, compared to using $SVM_{ta}^{2}$ for the issue reports without any attachments, using $SVM_{hybrid}$ (as it actually leverages the $SVM_{t}^{1})$ model for these reports) increased the accuracy from $0.848$ to $0.851$.

{\bf Regarding the statistical significance of the results.} We then analyzed whether the differences in the assignment accuracies obtained from the best performing multi-source models presented in this work, i.e., $SVM_{ta}^{2}$ and $SVM_{hybrid}$, and those obtained from the currently deployed model in the field, namely $SVM_{t}^{1}$, which also turns out to be the best performing single-source model in this work, are statistically significant. To this end, we carried out a series of experiments. 

In the first set of experiments, we used exactly the same test set with the experiments discussed above (i.e., all the issue reports submitted in August $2019$), but varied the training set by choosing a subset of all the issue reports submitted within the last $6$ months of August $2019$, such that the training and test sets correspond to the $80$\% and $20$\% of all of the issue reports selected, respectively. We, furthermore, repeated the experiments $30$ times.

\begin{figure}[h]
  \centering
  \includegraphics[width=\linewidth]{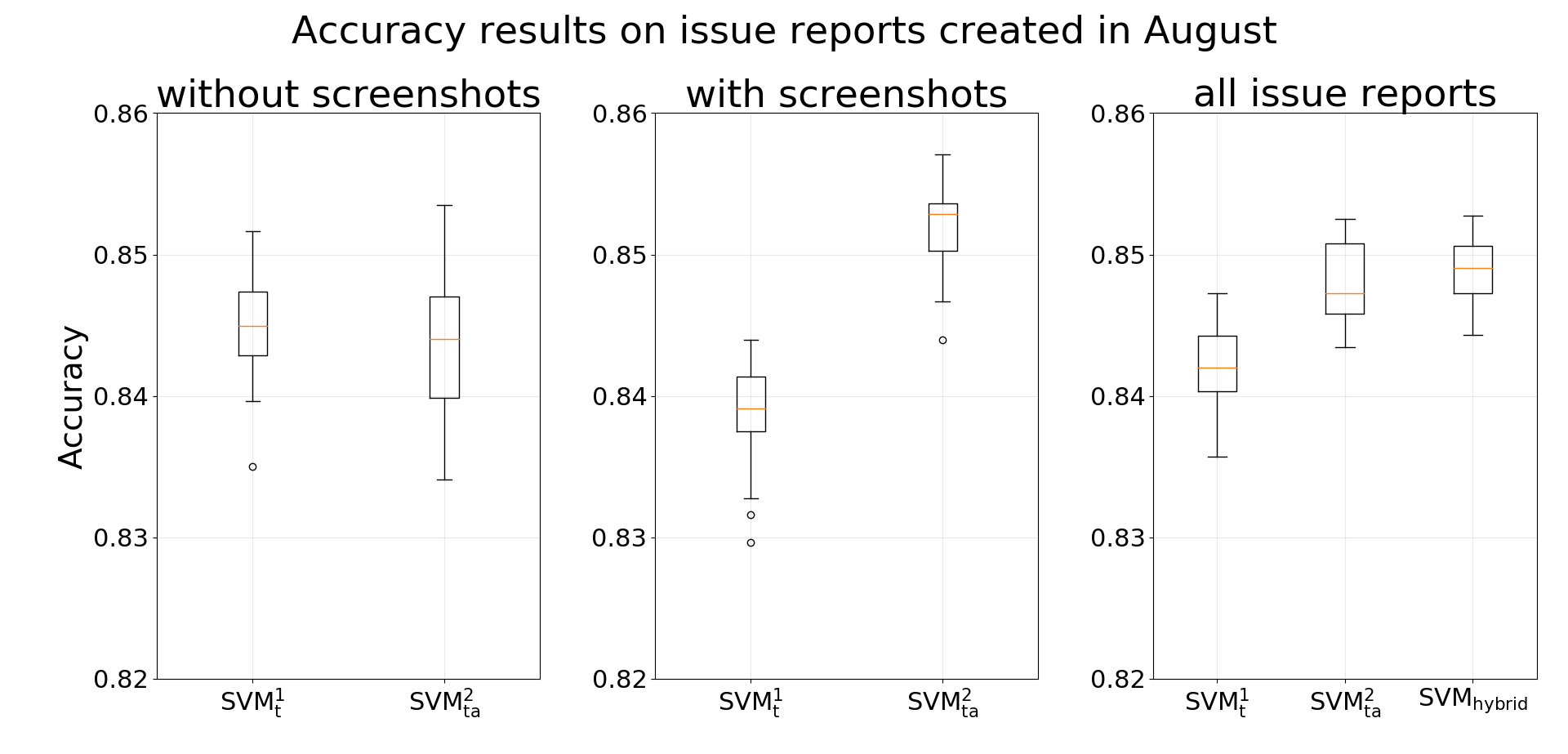}
  \caption{Box-whisker plots of the accuracies obtained on the issue reports submitted in August $2019$. The plots (from left right) present the distributions of the results obtained from the issue reports without and with attachments, and those obtained from all the issue reports in the test set, respectively. For each category, the experiments were repeated $30$ times.}
  \label{fig:box_plot_aug_data}
\end{figure}

Figure~\ref{fig:box_plot_aug_data} presents the distributions of the accuracies obtained from the $SVM_{t}^{1}$, $SVM_{ta}^{2}$, and $SVM_{hybrid}$ models for the issue reports with and without screenshot attachments as well as for all the issue reports in the test sets. Furthermore, while Table~\ref{tbl:stat_significance} presents the summary statistics for the results, Table~\ref{tbl:p_values} summarizes the results of the statistical significance tests where the entries in bold represent the statistically significant results (look for the month of August in both tables). Note that we report the results of $SVM_{hybrid}$ only for the entire test set as this model uses either the $SVM_{t}^{1}$ or the $SVM_{ta}^{2}$ model for the assignments, depending on the presence of the attachments.

\begin{table*}
\small
\caption{The summary statistics for the accuracy (A) and the measure (F) values obtained in different experimental setups. For each setup, the experiments were repeated $30$ times.}
\label{tbl:stat_significance}      
\resizebox{\columnwidth}{!}   
{\begin{tabular}{|l|l|rr|rr|rr|rr|rr|rr|rr|}
\hline
\multirow{3}{*}{month}   &  \multirow{3}{*}{stat} &  \multicolumn{4}{|c|}{test set w/o screenshots} & \multicolumn{4}{|c|}{test set w/ screenshots} &\multicolumn{6}{|c|}{all test set}  \\
\cline{3-16}
 & & \multicolumn{2}{|c|}{$SVM_{t}^{1}$} & \multicolumn{2}{|c|}{$SVM_{ta}^{2}$} & \multicolumn{2}{|c|}{$SVM_{t}^{1}$} & \multicolumn{2}{|c|}{$SVM_{ta}^{2}$} & \multicolumn{2}{|c|}{$SVM_{t}^{1}$} & \multicolumn{2}{|c|}{$SVM_{ta}^{2}$} & \multicolumn{2}{|c|}{$SVM_{hybrid}$} \\
 \cline{3-16}
 & & \multicolumn{1}{c}{A} & \multicolumn{1}{c|}{F} & \multicolumn{1}{c}{A} & \multicolumn{1}{c|}{F} & \multicolumn{1}{c}{A} & \multicolumn{1}{c|}{F} & \multicolumn{1}{c}{A} & \multicolumn{1}{c|}{F} & \multicolumn{1}{c}{A} & \multicolumn{1}{c|}{F} & \multicolumn{1}{c}{A} & \multicolumn{1}{c|}{F} & \multicolumn{1}{c}{A} & \multicolumn{1}{c|}{F} \\
\hline
\multirow{4}{*}{August} & mean & 0.845 & 0.836 & 0.844 & 0.834 & 0.839 & 0.826 & 0.852 & 0.839 & 0.842 & 0.831 & 0.848 & 0.837 & 0.849 & 0.837 \\
& std. & 0.004 & 0.004 & 0.005 & 0.005 & 0.004 & 0.004 & 0.003 & 0.003 & 0.003 & 0.003 & 0.003 & 0.003 & 0.002 & 0.002 \\
& max & 0.852 & 0.843 & 0.854 & 0.843 & 0.844 & 0.833 & 0.857 & 0.845 & 0.847 & 0.836 & 0.853 & 0.842 & 0.853 & 0.842 \\
& min & 0.835 & 0.826 & 0.834 & 0.821 & 0.830 & 0.817 & 0.844 & 0.829 & 0.836 & 0.825 & 0.844 & 0.832 & 0.844 & 0.833 \\
\hline
\multirow{4}{*}{July} & mean & 0.851 & 0.835 & 0.847 & 0.830 & 0.843 & 0.832 & 0.853 & 0.842 & 0.846 & 0.833 & 0.850 & 0.838 & 0.852 & 0.840 \\
& std. & 0.003 & 0.003 & 0.003 & 0.003 & 0.002 & 0.002 & 0.002 & 0.002 & 0.002 & 0.002 & 0.002 & 0.002 & 0.002 & 0.002 \\
& max & 0.856 & 0.839 & 0.851 & 0.835 & 0.849 & 0.837 & 0.858 & 0.847 & 0.850 & 0.836 & 0.855 & 0.842 & 0.856 & 0.842 \\
& min & 0.845 & 0.830 & 0.843 & 0.827 & 0.837 & 0.826 & 0.848 & 0.836 & 0.841 & 0.828 & 0.846 & 0.835 & 0.848 & 0.836 \\
\hline
\multirow{4}{*}{June}& mean & 0.831 & 0.822 & 0.828 & 0.819 & 0.835 & 0.820 & 0.841 & 0.826 & 0.834 & 0.822 & 0.835 & 0.824 & 0.836 & 0.825 \\
& std. & 0.004 & 0.004 & 0.004 & 0.004 & 0.003 & 0.003 & 0.003 & 0.003 & 0.003 & 0.002 & 0.002 & 0.002 & 0.002 & 0.002 \\
& max & 0.840 & 0.830 & 0.834 & 0.827 & 0.842 & 0.827 & 0.845 & 0.831 & 0.840 & 0.828 & 0.840 & 0.829 & 0.841 & 0.830 \\
& min & 0.823 & 0.816 & 0.818 & 0.810 & 0.830 & 0.815 & 0.835 & 0.821 & 0.828 & 0.817 & 0.831 & 0.821 & 0.832 & 0.821 \\
\hline
\multirow{4}{*}{May} & mean & 0.835 & 0.818 & 0.831 & 0.814 & 0.831 & 0.818 & 0.840 & 0.827 & 0.833 & 0.819 & 0.836 & 0.821 & 0.838 & 0.823 \\
& std. & 0.003 & 0.003 & 0.003 & 0.003 & 0.002 & 0.002 & 0.002 & 0.002 & 0.002 & 0.002 & 0.002 & 0.002 & 0.002 & 0.002 \\
& max & 0.839 & 0.823 & 0.836 & 0.818 & 0.835 & 0.822 & 0.844 & 0.831 & 0.837 & 0.822 & 0.839 & 0.824 & 0.840 & 0.826 \\
& min & 0.828 & 0.812 & 0.827 & 0.809 & 0.829 & 0.816 & 0.837 & 0.824 & 0.830 & 0.815 & 0.833 & 0.819 & 0.835 & 0.820 \\
\hline
\multirow{4}{*}{April} & mean & 0.834 & 0.814 & 0.831 & 0.809 & 0.813 & 0.800 & 0.828 & 0.813 & 0.823 & 0.806 & 0.828 & 0.811 & 0.830 & 0.813 \\
& std. & 0.003 & 0.003 & 0.003 & 0.003 & 0.002 & 0.002 & 0.002 & 0.002 & 0.002 & 0.002 & 0.002 & 0.002 & 0.002 & 0.002 \\
& max & 0.838 & 0.818 & 0.837 & 0.816 & 0.817 & 0.803 & 0.832 & 0.817 & 0.825 & 0.808 & 0.831 & 0.814 & 0.833 & 0.816 \\
& min & 0.827 & 0.808 & 0.824 & 0.802 & 0.808 & 0.794 & 0.824 & 0.810 & 0.820 & 0.802 & 0.825 & 0.808 & 0.827 & 0.810 \\
\hline
\multirow{4}{*}{March} & mean & 0.845 & 0.829 & 0.845 & 0.828 & 0.825 & 0.813 & 0.844 & 0.832 & 0.834 & 0.820 & 0.844 & 0.830 & 0.844 & 0.831 \\
& std. & 0.004 & 0.004 & 0.003 & 0.003 & 0.002 & 0.002 & 0.003 & 0.003 & 0.002 & 0.002 & 0.002 & 0.002 & 0.003 & 0.003 \\
& max & 0.851 & 0.835 & 0.850 & 0.833 & 0.828 & 0.816 & 0.849 & 0.839 & 0.839 & 0.825 & 0.847 & 0.833 & 0.849 & 0.836 \\
& min & 0.837 & 0.823 & 0.837 & 0.822 & 0.821 & 0.809 & 0.839 & 0.826 & 0.830 & 0.816 & 0.840 & 0.826 & 0.840 & 0.826 \\
\hline
\end{tabular}}
\end{table*}

\begin{table*}
\small
\caption{The results of the non-parametric Wilcoxon rank-sum tests. A p-value less than $0.05$ is considered to be statistically significant, which are presented in bold.}
\label{tbl:p_values}      
\resizebox{\columnwidth}{!}   
{\begin{tabular}{|llc|r|r|r|r|r|r|}
\hline
\multirow{2}{*}{model 1}&\multirow{2}{*}{model 2}&\multirow{2}{*}{test set}&
\multicolumn{6}{|c|}{p-values} \\
\cline{4-9}
 & & & \multicolumn{1}{|c|}{August} & \multicolumn{1}{|c|}{July} & \multicolumn{1}{|c|}{June}  & \multicolumn{1}{|c|}{May} & \multicolumn{1}{|c|}{April} & \multicolumn{1}{|c|}{March} \\
\hline
$SVM_{t}^{1}$ & $SVM_{ta}^{2}$ & w/ screenshots & \textbf{0.000001731} & \textbf{0.000001724} & \textbf{0.000004273} &  \textbf{0.000001732} & \textbf{0.000001727} & \textbf{0.000001731} \\
$SVM_{t}^{1}$ & $SVM_{ta}^{2}$ & w/o screenshots & 0.06259948 & \textbf{0.000001726} & \textbf{0.000259608} &  \textbf{0.000001721} & \textbf{0.000006941} & \textbf{0.000001727} \\
$SVM_{t}^{1}$ & $SVM_{ta}^{2}$ & all & \textbf{0.000001725} & \textbf{0.000001732} & \textbf{0.001952822} &  \textbf{0.000005175} & \textbf{0.000001733} & \textbf{0.039670807} \\
$SVM_{t}^{1}$ & $SVM_{hybrid}$ & all & \textbf{0.000001726} & \textbf{0.000001733} & \textbf{0.000031003} &  \textbf{0.000001729} & \textbf{0.000001730} & \textbf{0.000001719} \\
$SVM_{ta}^{2}$ & $SVM_{hybrid}$ & all & \textbf{0.016215804} & \textbf{0.000002841} & \textbf{0.006275772} &  \textbf{0.000001725} & \textbf{0.000002544} & 0.592647639 \\
\hline
\end{tabular}}
\end{table*}

We observed exactly the same trends with our original experiments. In particular, $SVM_{t}^{1}$ generally performed better than $SVM_{ta}^{2}$ for the issue reports without any attachments; an average accuracy of $0.845$ vs. $0.844$. For the issue reports with attachments, on the other hand, $SVM_{ta}^{2}$ performed profoundly better than $SVM_{t}^{1}$; an average accuracy of $0.852$ vs. $0.839$. The difference was indeed statistically significant (Table~\ref{tbl:p_values}). 

Overall, i.e., when all the issue reports with and without screenshot attachments are taken into account, $SVM_{hybrid}$ performed significantly  better (both in the practical and in the statistical sense) than the currently deployed model in the field (i.e., $SVM_{t}^{1}$); an average accuracy of $0.849$ vs. $0.842$ (Table~\ref{tbl:p_values}).

In the second set of experiments, we have also varied the test sets. In particular, we repeated experiments we carried out for August for each remaining month $m$ from March to July (inclusive) by using all the data submitted in the month of $m$ as the test set and by randomly picking a training set from the issue reports submitted within the last $6$ months of $m$, such that the training and test sets represent $80$\% and $20$\% of all the issue reports selected, respectively. The experiments were again repeated $30$ times for each experimental setup. For this set of experiments, we used an additional set of $43,930$ distinct issue reports (Table~\ref{tab:data}).

\begin{figure*}
\centering
  \subfloat[]{
    \includegraphics[width=\textwidth,height=0.2\textheight]{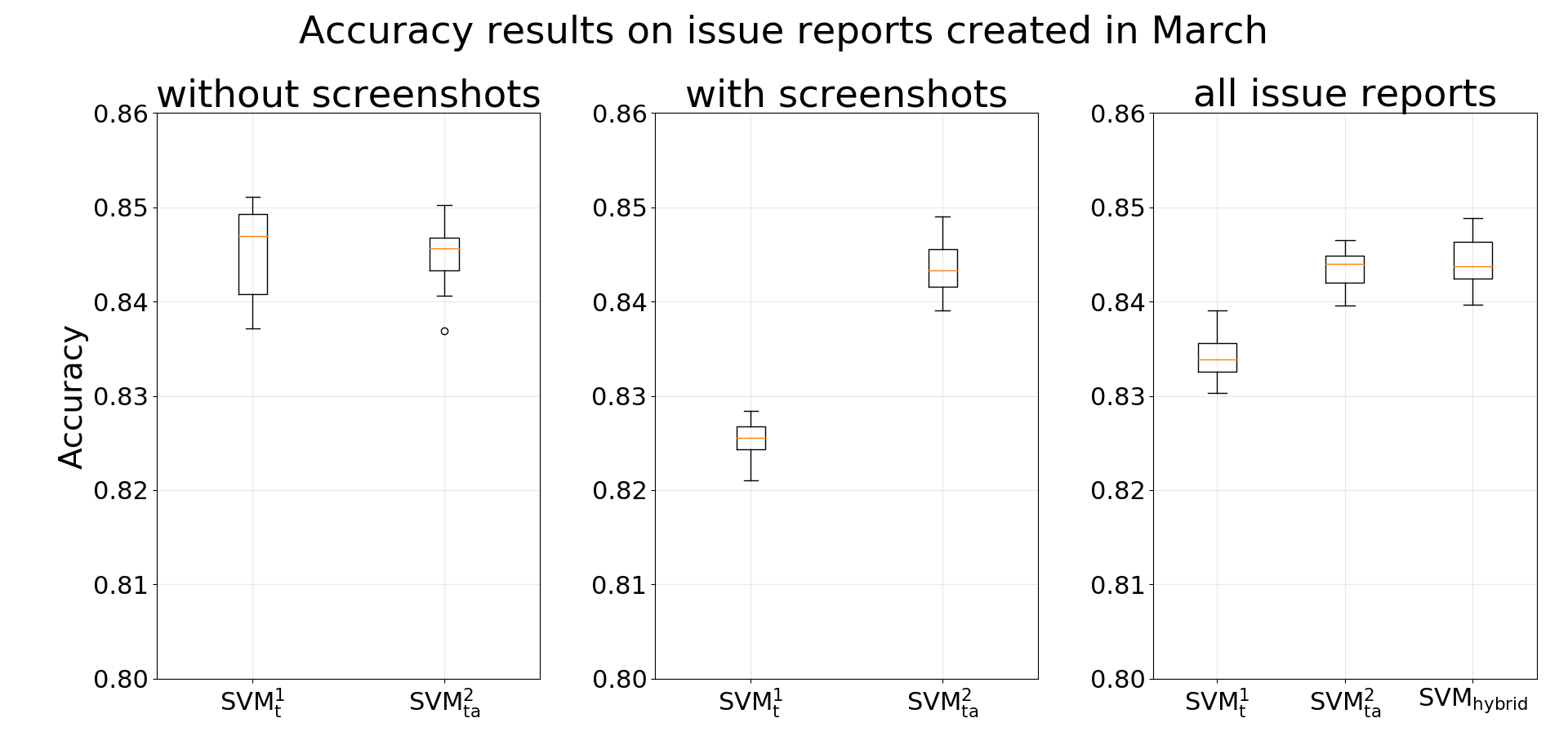}}\hfill
  \subfloat[]{
    \includegraphics[width=\textwidth,height=0.2\textheight]{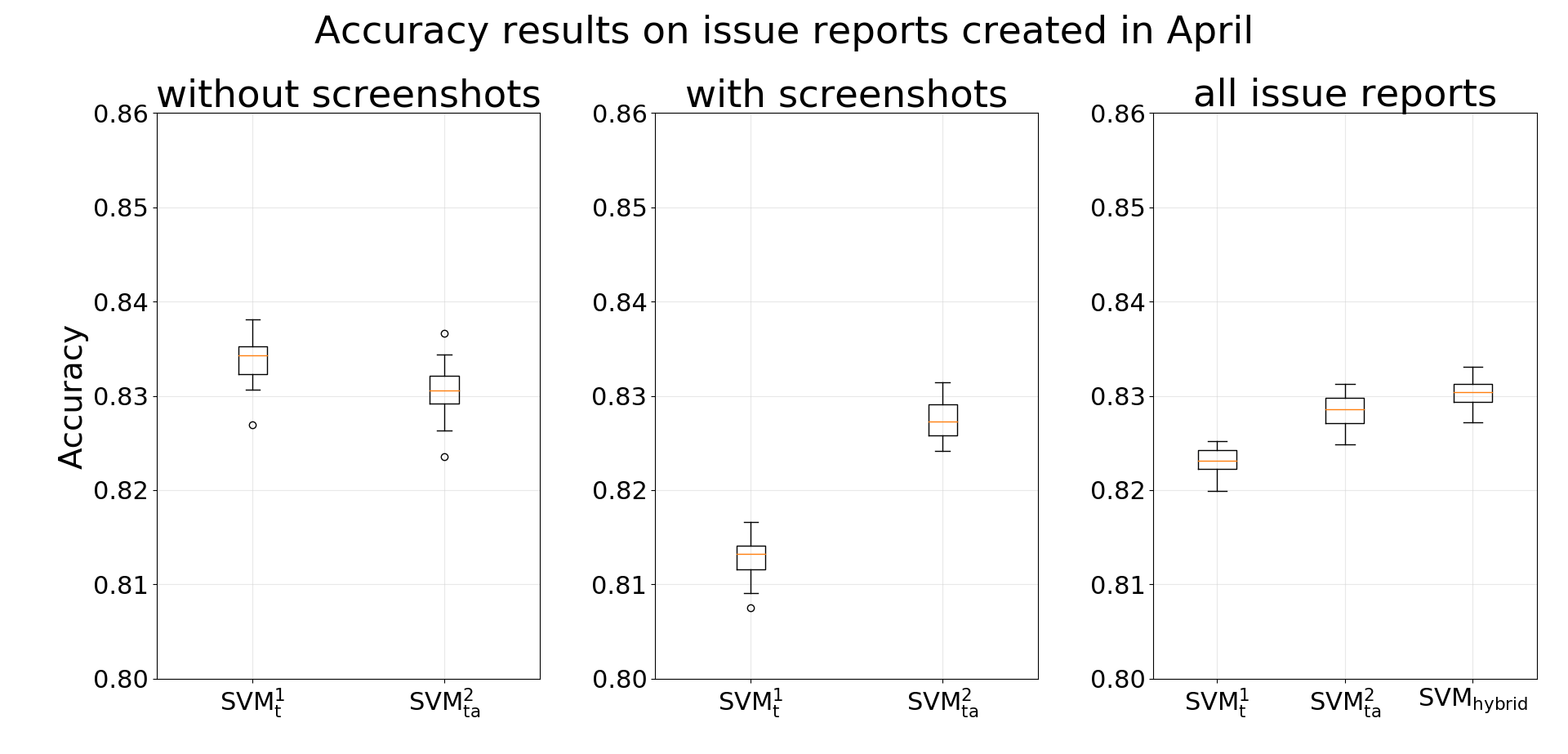}}\hfill
  \subfloat[]{
    \includegraphics[width=\textwidth,height=0.2\textheight]{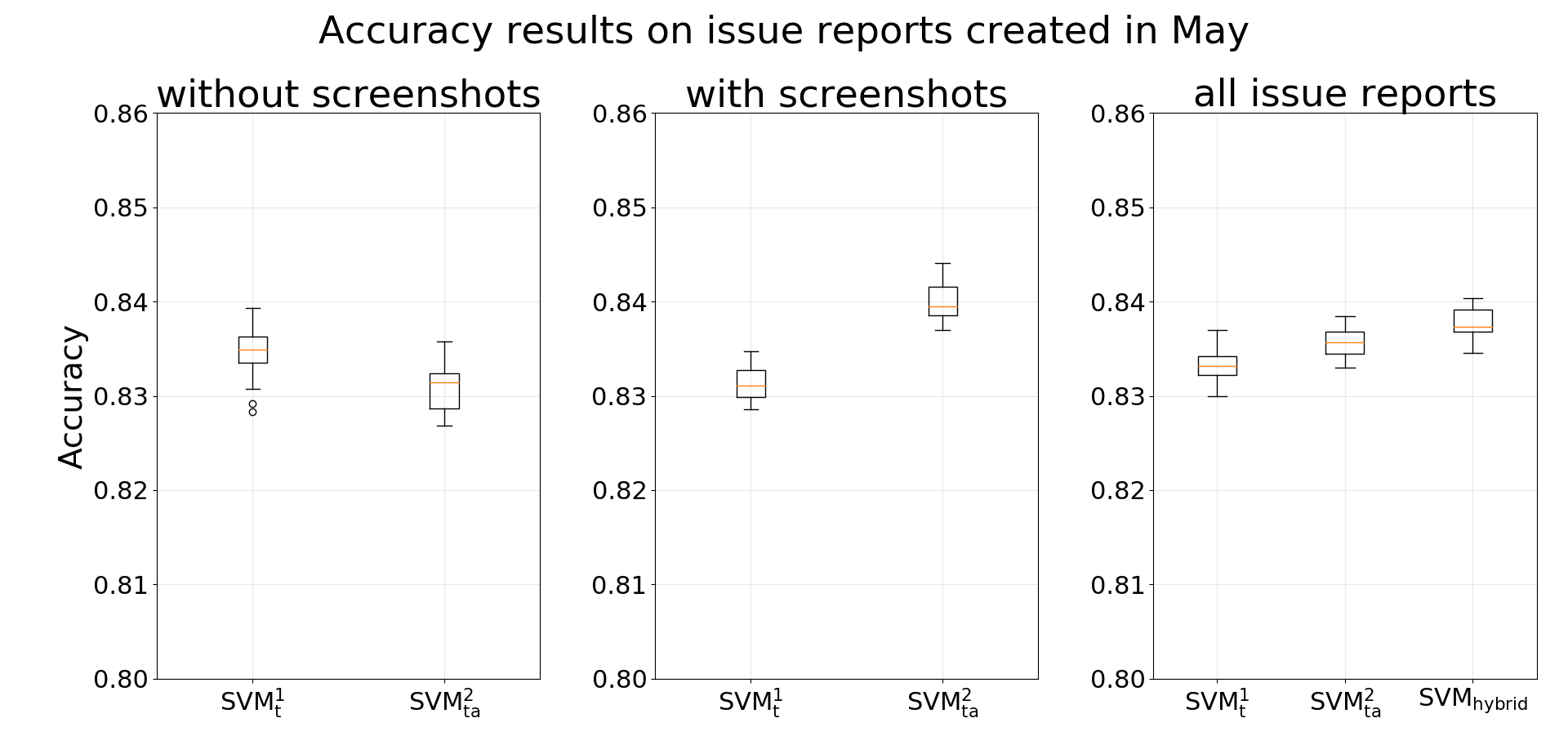}}\hfill
  \subfloat[]{
    \includegraphics[width=\textwidth,height=0.2\textheight]{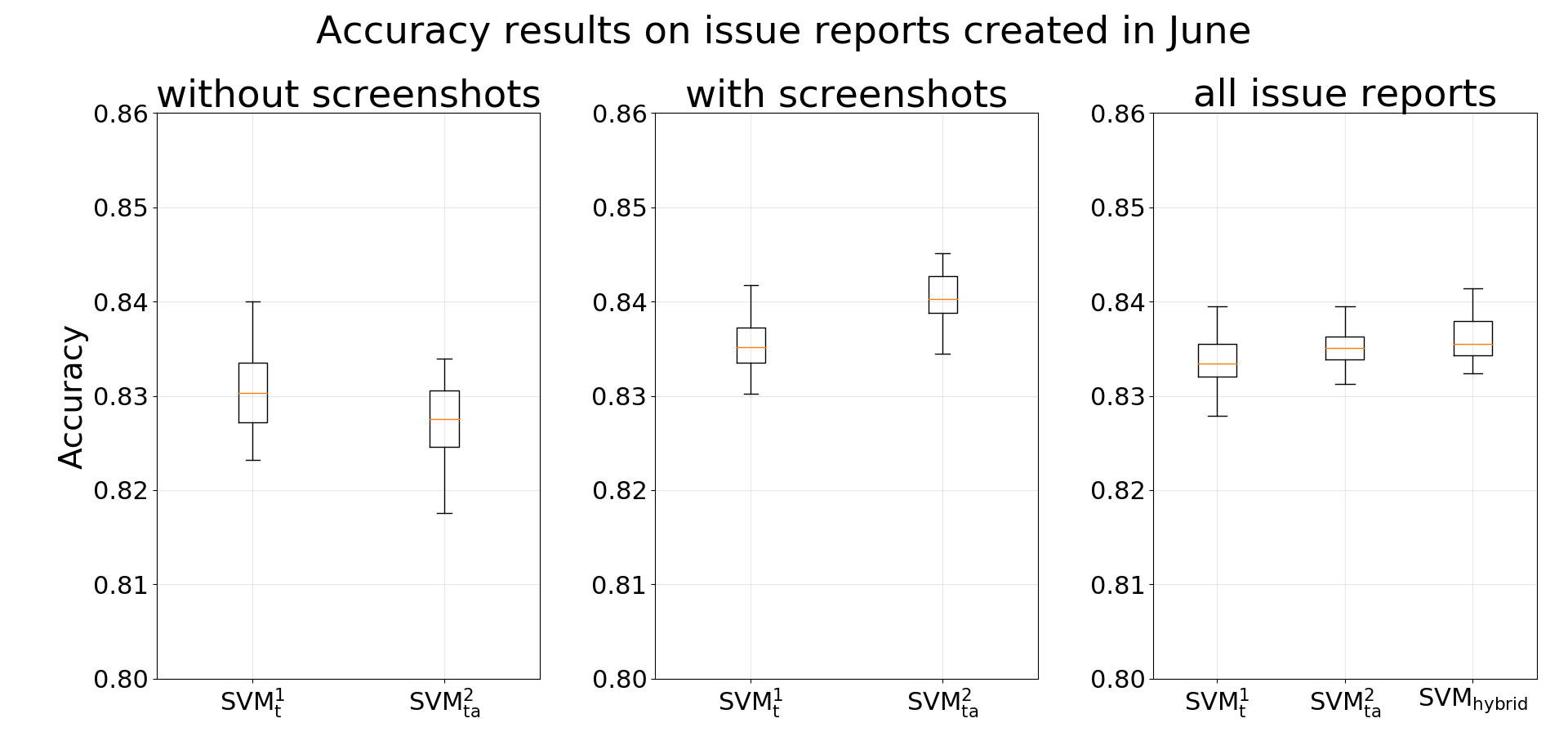}}\hfill
  \subfloat[]{
    \includegraphics[width=\textwidth,height=0.2\textheight]{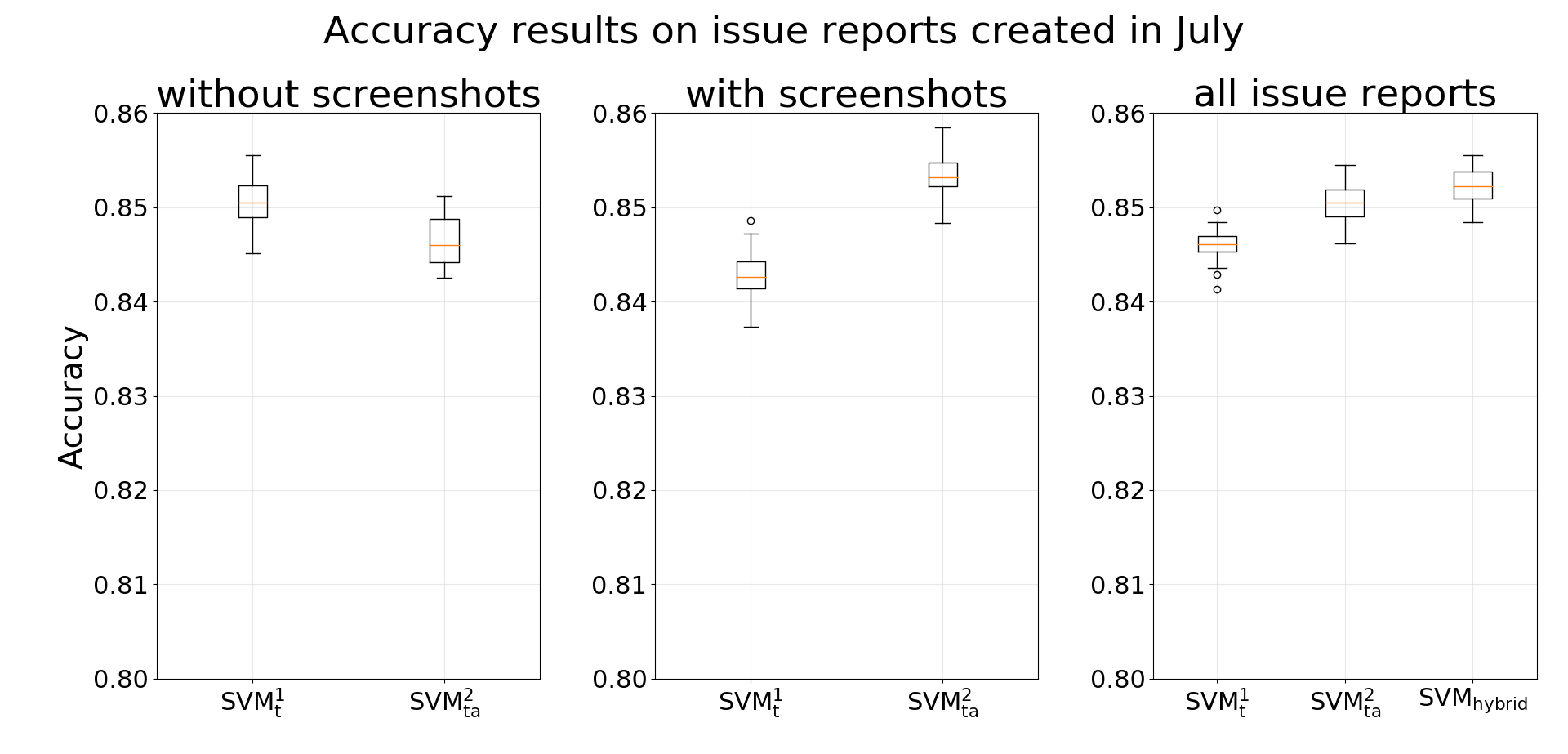}}\hfill
  \caption{ Box-whisker plots of the accuracies obtained for the issue reports submitted in a) March, b) April, c) May, d) June, and e) July of $2019$. For each category (i.e., for each box plot), the experiments were repeated $30$ times.}
\label{fig:box_plot_all_data}
\end{figure*}

Figure~\ref{fig:box_plot_all_data} and Table~\ref{tbl:stat_significance} present the results we have obtained. We observed exactly the same trends with the previous set of experiments: 1) $SVM_{t}^{1}$ generally performed better than $SVM_{ta}^{2}$ for the issue reports without any screenshot attachments; 2) for the ones with screenshot attachments, however, $SVM_{ta}^{2}$ performed profoundly better than $SVM_{t}^{1}$; and 3) overall, $SVM_{hybrid}$ was the best performing model.

Furthermore, the statistical significance tests revealed that almost all the differences between the aforementioned models (except for the difference between the $SVM_{ta}^2$ and $SVM_{hybrid}$ models for the month of March) were indeed statistically significant in each experimental setup (Table~\ref{tbl:p_values}). Note that these results not only support the claims of the paper, but also further justify the need for the $SVM_{t}^{1}$, $SVM_{ta}^{2}$, and $SVM_{hybrid}$ models.

{\bf Regarding RQ3: How does taking the screenshot attachments into account affect the overall performance in terms of the training and the prediction times?} To address this research question, we compared the performance of our best performing multi-source model (i.e., $SVM_{ta}^{2}$) with that of the deployed model in the field (i.e., $SVM_{t}^{1}$). All the experiments were carried out on a Dual-Core Intel(R) Core(TM) i7-6600U CPU @2.60 GHz computer with 16GB of RAM running Windows 10 Enterprise 2017 as the operating system. 

In particular, we used three metrics for the comparisons; OCR time, response time, and training time. The first metric measures the end-to-end processing time required for extracting the text from a single screenshot attachment, including the time required for loading the attachment to the memory. We report this metric on a per screenshot attachment basis as this step can easily be parallelized; multiple attachments can be processed in parallel. The second metric, i.e., the response time of a model, is measured as the end-to-end time required for assigning a given issue report. Note that the response times also include the OCR times (for $SVM_{ta}^{2}$), because the text in a given screenshot attachment needs to be extracted before the assignment can be made. And, the third metric, i.e., the training time of a model, is measured as the time it takes to train the model once all the data to flow through the channels is fed as input.  That is, OCR times are not included in the training times. The reason for this is two folds. First, the model employed by IssueTAG is re-trained as needed~\citep{aktas2020automated}. Therefore, the texts extracted from the screenshot attachments for the purpose of assigning the respective issue reports to the development teams, can be saved to re-train the model in the future. That is, the OCR step needs to be carried out only once during the assignment, which is, indeed, accounted for in the response times. Second, as discussed before, the OCR step can easily be parallelized (e.g., for the training of the very first model).


Table \ref{tbl:timeperf} presents the time measurements (in seconds) we obtained by repeating each experiment at least $30$ times. We observed that, although taking the screenshot attachments into account when assigning the issue reports, expectedly increased the training and the response times, all of these overheads were acceptable at Softtech. More specifically, the $SVM_{ta}^{2}$ model, compared to the $SVM_{t}^{1}$ model, increased the average training time from $190.4$ to $317.2$ seconds and the average response time from $0.9$ to $2.17$ seconds. A significant portion of the response time for the $SVM_{ta}^{2}$ model ($2.11$ out $2.17$) was indeed spent for OCR.

\begin{table}
\caption{Running times (in seconds) of various operations.}
\label{tbl:timeperf}
\centering 
\begin{tabular}{|l|r|r|r|r|r|r|}
\hline
metric  & mean & std. & max & min & median \\ 
\hline  
    OCR time & 2.11 & 1.05 & 8.51 & 0.57 & 1.85 \\
    Training time for $SVM_{t}^{1}$ & 190.4 & 6.69 & 202 & 180 & 190.5 \\
    Training time for $SVM_{ta}^{2}$ & 317.2 & 17.08 & 348 & 291 & 314.5 \\
    Response time for $SVM_{t}^{1}$ & 0.9 & 0.03 & 1.01 & 0.86 & 0.9 \\
    Response time for $SVM_{ta}^{2}$ & 2.17 & 0.09 & 2.54 & 2.07 & 2.16 \\
    \hline
\end{tabular}
\end{table}

Note further that, since the $SVM_{hybrid}$ model requires both the $SVM_{t}^{1}$ and the  $SVM_{ta}^{2}$ models to be trained, the training time for this model will be $507.6$ seconds (the sum of the training times for the required models). And, the response time of the model will be $0.9$ seconds for the issue reports without any screenshot attachments (as the $SVM_{t}^{1}$ model is used) and $2.17$ seconds for the ones with the screenshot attachments (as the $SVM_{ta}^{2}$ model is used), on average.

\section{Threats to Validity}
\label{threats}


\subsection{Construct Validity}
\label{constructValidity}

To circumvent the construct threats, we used the well-known accuracy metric together with the other frequently used metrics, namely precision, recall, and F-measure~\citep{murphy2004automatic, anvik2006should, baysal2009bug, anvik2011reducing, jeong2009improving, bhattacharya2012automated, jonsson2016automated,  dedik2016automated, manning2008introduction}. The discussions in the paper mainly focused on the accuracy results as this metric has been the choice of discussion in some of the recent related works~\citep{aktas2020automated, jonsson2016automated}. After all, as also reported in the paper, the remaining metrics exhibited the same (or similar) trends with the accuracy metric. 

We, furthermore, measured the cost for different models in terms of the amount of time required to carry out the integral tasks regarding both the construction and the uses of the models. We did this because the running times of the proposed approaches were the most important concern at Softtech.

\subsection{Internal Validity}
\label{internalValidity}

To alleviate the threats to internal validity, we used mature tools to carry out the integral computations required by the proposed approaches. More specifically, we used py-tesseract~\citep{smith2007overview} for OCR; scikit-learn~\citep{pedregosa2011scikit} for tf-idf vectorization and linear SVM classification; keras~\citep{chollet2015keras} for word embeddings and CNN classification; and PyTorch deep learning framework~\citep{paszke2019pytorch} for the multimodal model. 

We have, furthermore, employed well-known and frequently-used pre-processing steps to analyze the text extracted from the issue reports and the screenshot attachments, including tokenization and removal of non-letter characters~\citep{manning2008introduction}. Similarly, the architectures of the machine learning models we used to extract and analyze the visual features present in the screenshot attachments, namely $VGG_{\hat{a}}^1$ and $MML_{t\hat{a}}^2$, were also published in the literature~\citep{simonyan2014very, joulin2016bag, bojanowski2017enriching, he2016deep}.

We (unless otherwise stated) used the machine learning models with their default configurations. The performance of these models in the experiments might have been dependent on the underlying  configurations. Note, however, that optimizing the configurations could have only improved the accuracy of the models.

Last but not least, we have checked the validity of the results manually by using manageable-size test sets. We also repeated the experiments by using different collections of training and test sets and observed the same trends.

\subsection{External Validity}
\label{externalValidity}

One external threat is that all of the issue reports used in the study were submitted to only one company, namely Softtech. However, Softtech is the largest software development company owned by domestic capital in Turkey. As Softtech produces and maintains dozens of business-critical systems comprised of hundreds of millions of lines of code, it shares many characteristics with the vendors of other business-critical systems, such as developing custom software systems; having a large, evolving codebase maintained by dozens of development teams; and receiving a large number of issue reports from the field, each of which generally needs to be addressed with utmost importance and urgency.

Another threat is that the development teams at Softtech typically use a small number of UI frameworks with a quite strict guidelines for designing the user interfaces. This makes the products produced by different teams to have the same/similar look and feel, which we believe was the main reason as to why the visual features extracted from the screenshot attachments were not helpful at all in the assignments. Therefore, we believe that, in scenarios where the look and feel of the products varies depending on the development teams, the visual features can still play an important role in the assignments.

\subsection{Conclusion Validity}
\label{conclusionValidity}

All the issue reports used in this work were real issue reports submitted to Softtech. Furthermore, the period of time selected for the study was representative of the issue report database maintained by Softtech, in terms of the number of issue reports submitted, the percentage of the issue reports with screenshot attachments, and the number of development teams, to which these issue reports are assigned.

Furthermore, we used only the issue reports, which were marked as closed with the ``resolved'' status, in order not to introduce any bias in the assignment accuracies. At Softtech, the issue reports are closed by the development teams, who resolve the reported issues. Since the number of issue reports resolved by a team is used as a key performance indicator at Softtech, the developers pay utmost attention to correctly indicate the teams closing the issue reports. For a given issue report, we, therefore, used the development team, who closed the report, as the ground truth.

\section{Related Work}
\label{relatedWork}

Murphy et al. report that as the software systems are getting bigger, the issue triaging takes increasingly larger amount time~\citep{murphy2004automatic}. Therefore, many approaches have been proposed in the literature to automate the process of issue triaging~\citep{ahsan2009automatic, alenezi2013efficient, anvik2011reducing, podgurski2003automated, anvik2006should, baysal2009bug, jeong2009improving, bhattacharya2012automated, lin2009empirical, helming2010automatic, park2011costriage, xia2013accurate, xie2012dretom, dedik2016automated, jonsson2016automated, lee2017applying, chen2019empirical, gu2020efficient, zhang2020efficient, sajedi2020vocabulary, aung2021multi, chmielowski2021impact}. 

Many of these approaches use only the one-line summaries and/or descriptions of the issue reports to assign them to the stakeholders for resolutions. There are, however, approaches that utilize different sources of additional information to further improve the assignment accuracy, including the contents of the duplicated issue reports~\citep{bettenburg2008duplicate}; human triagers by offering them an automatically generated, ranked list of recommended assignees, so that they can select the assignee~\citep{anvik2006should}; tossing histories of the issue reports~\citep{jeong2009improving}; the expertise of the developers (inferred from the attributes of the software products/components they modify to resolve the previously reported issue reports)~\citep{bhattacharya2012automated}; additional features extracted from the issue reports, such as the priority, the submitter, the affiliation of the submitter, the site from where the report was submitted, and the version of the product, for which the issue report was submitted~\citep{lin2009empirical, jonsson2016automated}; and the relationships between the issue reports resolved by stakeholders and the respective functional requirements for the assignments~\citep{helming2010automatic}. Our work differ from these works in that we use the screenshot attachments in the issue reports as an additional source of information for issue assignment.


Some recent works focus on using video recordings as issue reports~\citep{cooper2021tango, bernal2020translating}. More specifically, \cite{bernal2020translating} use video recordings to automatically reproduce the reported issues and \cite{cooper2021tango} use them to identify duplicated reports. These approaches, however, mainly target mobile platforms and their applicability to the other computing platforms, especially for the stateful applications, the behaviors of which depend on the often persisted states (such as the ones stored in databases), is still an open question. For example, in the real issue reports submitted to Softtech, we did not have any screen recordings as attachments. Furthermore, the assignment problem, which is the main focus of this work, has not been addressed by the aforementioned works. Last but not least, our work uses screenshot attachments as an additional source of information for the assignments, not as the only source of information. However, using screen recordings in a similar manner for issue assignment is certainly an interesting avenue for future research.

While many of the existing works were evaluated on open source projects~\citep{murphy2004automatic, anvik2006should, baysal2009bug, ahsan2009automatic, jeong2009improving, anvik2011reducing, bhattacharya2012automated, park2011costriage, alenezi2013efficient, xia2013accurate, xie2012dretom, sajedi2020vocabulary, aung2021multi}, few were evaluated on commercial, closed-source projects~\citep{dedik2016automated, helming2010automatic, jonsson2016automated, lee2017applying, lin2009empirical, chen2019empirical, gu2020efficient, zhang2020efficient, chmielowski2021impact, oliveira2021issue}. Compared to the former set of works, we evaluate the proposed approach in a large industrial setup where hundreds of millions of lines of mostly business-critical codes were maintained by dozens of development teams. Compared to the latter set of works, our system, IssueTAG, is actually deployed in the field, automatically assigning all the issue reports submitted since it is deployment on $January$ $2018$ ($301,752$ issue reports as of $November$ $2021$). We have indeed been constantly maintaining IssueTAG by fine tuning it to further improve the assignment accuracy and by enhancing it with additional features. 

One difference we observe when it comes to the issue reports submitted to open source projects and the ones submitted to Softtech is that, the former tend to have more technical information, including the information regarding the internal workings of the systems. The latter, on the other hand, typically describe the symptoms of the failures as they are observed from outside the system by non-technical end users. Furthermore, the latter set of issue reports tend to be written more formally with little or no language errors at all, whereas the former set of issue reports tend to be written informally with grammar mistakes and typos.

Many of the existing works prefer to assign the issue reports to the individual developers~\citep{ahsan2009automatic, alenezi2013efficient, anvik2006should, baysal2009bug, bhattacharya2012automated, jeong2009improving, murphy2004automatic, park2011costriage, xia2013accurate, xie2012dretom, sajedi2020vocabulary, aung2021multi}. In this work, however, we assign them to the development teams as with~\citep{jonsson2016automated, chmielowski2021impact}. This, which was also the case before the deployment of IssueTAG, is indeed a decision deliberately made by Softtech to take the team dynamics into account during the assignments, which are quite difficult to model, such as the current workloads of the individual developers, the changes in the team structures, and the current status of the developers. After all many of the development teams at Softtech are close-knit teams following agile development processes. Note further that since the assignees are modeled as classes, there is no theoretical limit to the application of the proposed models for assigning the issue reports to the individual developers.

There are also other issue triaging-related tasks, including the identification of duplicated bug reports~\citep{runeson2007detection, bettenburg2008duplicate, cooper2021tango}; determination of the severity levels for the reported issues~\citep{menzies2008automated, lamkanfi2010predicting}; estimation of the effort required for resolving the reported issues~\citep{weiss2007long, giger2010predicting, zhang2013predicting}; separation of the issue reports indicating defects from the ones not indicating any defects~\citep{antoniol2008bug}; and the identification of the missing information in the issue reports~\citep{bettenburg2008makes}. We believe that in all these tasks leveraging the screenshot attachments as an additional source of information can improve the performance of the proposed approaches.

\section{Conclusion}
\label{conclusion}

In this work, we presented a number of approaches to use the screenshot attachments present in issue reports as an additional source of information for automated assignment. We, furthermore, evaluated all of the proposed approaches empirically by using a total of $84,972$ real issue reports submitted to Softtech.

The results of experiments strongly support our basic hypothesis that using screenshot attachments can further improve the assignment accuracy. We have arrived at this conclusion by noting that 1) a large fraction of all the issue reports submitted to Softtech has screenshot attachments; 2) in the presence of screenshot attachments, the one-line summary and the description fields of the issue reports often contain less information, compared to the issue reports without any attachments, which tend to reduce the assignment accuracy; 3) the screenshot attachments, on the other hand, convey invaluable information towards having better assignments; 4) for the issue reports with screenshot attachments, the assignment models, which use both sources of information (i.e., both the textual information present in the issue reports and the screenshot attachments) provided significantly (both in the practical and statistical sense) better accuracies compared to the models using a single source of information (i.e., either the textual information present in the issue reports or the screenshot attachments); and 5) all of these improvements were obtained at acceptable costs.

One potential avenue for future research is to further improve the assignment accuracy by focusing on the ``important'' regions in a given screenshot to filter out the parts, which create superficial commonalities between the issue reports. Another avenue is to use not only the screenshots, but also the other types of attachments for the assignments. Last but not least, the attachments can also be leveraged in other types of bug triaging-related analyses, including the determination of the severity levels, identification of the duplicates, and the estimation of the efforts required for resolving the reported issues.


%
%

\end{document}